\documentclass[a4paper,preprint,nofootinbib,11pt]{article}
\pdfoutput=1 

\usepackage{jheppub} 

\usepackage[T1]{fontenc} 
\usepackage[utf8]{inputenc}
\usepackage{mathrsfs,bm}
\usepackage{bibentry,natbib}
\usepackage{txfonts}
\usepackage{amssymb}
\usepackage{rotating}
\usepackage{indentfirst}
\usepackage{graphicx,booktabs}
\usepackage{multirow}
\usepackage{slashed}
\usepackage{overpic}
\usepackage{color}
\usepackage{amssymb}
\usepackage{hyperref}
\usepackage{bbding}
\usepackage{url}
\usepackage{ulem}
\allowdisplaybreaks

\title{New physics in $s\to d$ semileptonic transitions: rare hyperon vs. kaon decays}

\author[a,b,c]{Li-Sheng Geng}
\author[d,e]{Jorge Martin Camalich}
\author[f,a,1]{and Rui-Xiang Shi\note{Corresponding author.}}

\affiliation[a]{School of Physics, Beihang University, Beijing 102206, China}
\affiliation[b]{Beijing Key Laboratory of Advanced Nuclear Materials and Physics, Beihang University, Beijing, 102206, China}
\affiliation[c]{School of Physics and Microelectronics, Zhengzhou University, Zhengzhou, Henan 450001, China}
\affiliation[d]{Instituto de Astrof\'isica de Canarias, C/ V\'ia L\'actea, s/n E38205 - La Laguna, Tenerife, Spain}
\affiliation[e]{Universidad de La Laguna, Departamento de Astrof\'isica, La Laguna, Tenerife, Spain}
\affiliation[f]{School of Space and Environment, Beihang University, Beijing 102206, China}

\emailAdd{lisheng.geng@buaa.edu.cn}
\emailAdd{jcamalich@iac.es}
\emailAdd{ruixiang.shi@buaa.edu.cn}

\abstract{We investigate the potential of rare hyperon decays to probe the short distance structure in the $s\to d\nu\bar\nu$ and $s\to d\ell^+\ell^-$ transitions. Hyperon decays into neutrinos ($B_1\to B_2\nu\bar\nu$) can be reliably predicted by using form factors determined in baryon chiral perturbation theory. Their decay rates are sensitive to different short-distance operators, as compared to their kaon counterparts, and the corresponding branching fractions are in the range of $10^{-14}\sim10^{-13}$ in the standard model. In the context of the low-energy effective theory, we find that the anticipated BESIII measurements of the $B_1\to B_2\nu\bar\nu$ decays would lead to constraints on new physics in the purely axial vector $\bar d \gamma_\mu\gamma_5 s$ current that are stronger than the present limits from their kaon siblings $K\to \pi\pi\nu\bar\nu$. On the other hand, although hyperon decays into charged leptons are dominated by long-distance hadronic contributions, angular observable such as the leptonic forward-backward asymmetry is sensitive to the interference between long- and short-distance contributions. We discuss the sensitivity to new physics of a potential measurement of this observable in comparison with observables in the kaon decays $K_L\to\mu^+\mu^-$ and $K^+\to\pi^+\mu^+\mu^-$. We conclude that the current kaon bounds are a few orders of magnitude better than those that could be obtained from $\Sigma^+\to p\mu^+\mu^-$ except for two  scenarios with new physics in the $(\bar d \gamma^\mu s)(\bar\ell\gamma_\mu\gamma_5\ell)$ and $(\bar d \gamma^\mu \gamma_5s)(\bar\ell\gamma_\mu\ell)$ currents. Finally, we point out that the loop effects from renormalization group evolution are important in this context, when relating the low-energy effective field theory to new physics models in the UV.}

\begin{document}
\maketitle
\flushbottom

\section{Introduction}

The flavor-changing neutral current ~(FCNC) transitions $s\to d\nu\bar\nu$ and $s\to d\ell^+\ell^-$ are loop-induced and suppressed by the GIM mechanism~\cite{Glashow:1970gm} in the standard model (SM). Therefore they are ideal processes to test the SM and search for new physics (NP). Considerable effort has been invested in the kaon decays induced by these transitions (see Refs.~\cite{Buchalla:1998ba,Cirigliano:2011ny} for comprehensive reviews).
Recently, the neutrino modes $K^+\to\pi^+\nu\bar{\nu}$ and $K_L\to\pi^0\nu\bar{\nu}$ were measured by the
NA62 Collaboration~\cite{,NA62:2020fhy,NA62:2021zjw} and KOTO Collaboration~\cite{KOTO:2018dsc,KOTO:2020prk}. The best upper limits of the corresponding branching fractions at 90\% confidence level (CL) are, respectively:
\begin{eqnarray}
&&{\rm BR}(K^+\to\pi^+\nu\bar{\nu})<1.78\times10^{-10},\label{Kptopipvvexp}\\
&&{\rm BR}(K_L\to\pi^0\nu\bar{\nu})<3.0\times10^{-9}.
\end{eqnarray}
Both measurements  show a slight excess with respect to the SM expectations $\sim10^{-11}$~(see Sec.~III.B and the most recent predictions~\cite{Buras:2015qea,Brod:2021hsj}), which leaves little room for NP from vectorial couplings in the $s\to d$ currents.  On the other hand, the experimental  upper bounds of $K^+\to\pi^+\pi^0\nu\bar{\nu}$~\cite{E787:2000iwe} and $K_L\to\pi^0\pi^0\nu\bar{\nu}$~\cite{E391a:2011aa} at 90\% CL,
\begin{eqnarray}
&&{\rm BR}(K^+\to\pi^+\pi^0\nu\bar{\nu})<4.3\times10^{-5},\\
&&{\rm BR}(K_L\to\pi^0\pi^0\nu\bar{\nu})<8.1\times10^{-7},
\end{eqnarray}
are still very far from the SM predictions $\sim10^{-15}-10^{-13}$~(see Sec.~III.B and Refs.~\cite{Littenberg:1995zy,Chiang:2000bg,Geng:2020seh}). Thus, these decays provide only loose constraints on NP from axial-vectorial couplings which are not directly accessible in $K\to\pi\nu\bar{\nu}$.
In addition, the charged-lepton modes $K_L\to\mu^+\mu^-$ and $K^+\to\pi^+\mu^+\mu^-$, whose decay rates are dominated by long-distance contributions, cannot probe all the possible NP interactions in the $s\to d\ell^+\ell^-$ decays.
Therefore, it would be important to find other probes and measurements that allow us to thoroughly search for NP in these strangeness-changing transitions.

Hyperons, which carry net baryon number one and half-integer spin, can lead to substantially different decay modes, observables, and sensitivity to the underlying structure of the $s\to d\nu\bar\nu$ and $s\to d\ell^+\ell^-$ transitions. On the experimental side, the BESIII Collaboration has started a campaign of measurements of the $B_1\to B_2\nu\bar{\nu}$ decays by collecting  hyperon-antihyperon resonant  pair production at the $J/\psi$ peak~\cite{Li:2016tlt,BESIII:2018cnd,BESIII:2019nep}. Thus, this experiment can be run as a ``hyperon factory'' in the characteristically clean environment of an $e^+e^-$ collider, enabling the measurement of several semileptonic rare decays. In addition, the LHCb Collaboration has also started an ambitious program with the measurement of the branching  fraction of the  $\Sigma^+\to p\mu^+\mu^-$ decay~\cite{LHCb:2017rdd}(after first evidence reported by the HyperCP Collaboration~\cite{HyperCP:2005mvo}), which is expected to be followed by measurements of angular asymmetries after Run 2 of the LHC~\cite{AlvesJunior:2018ldo,Cerri:2018ypt}.

Until quite recently, little theoretical attention has been paid to rare hyperon decays. In a series of works~\cite{Tandean:2019tkm,He:2005yn,He:2018yzu,Su:2019tjn,Li:2019cbk,Hu:2018luj}, the authors have studied the rare hyperon decays $B_1\to B_2\nu\bar{\nu}$ and $\Sigma^+\to p\mu^+\mu^-$. However, only in the decays with missing energy these have been compared to the corresponding kaon modes~\cite{Tandean:2019tkm,Su:2019tjn,Li:2019cbk}.

In this work we systematically investigate the potential of rare hyperon decays to probe the $s\to d$ short-distance structure in both  neutrino and charged-lepton modes, putting special emphasis in the complementarity with the corresponding decays of their kaon siblings. In comparison with previous works, we also improve the nonperturbative inputs required for the form factors entering these decays by using model-independent methods such as isospin symmetry in $K\to\pi\pi\nu\bar\nu$ or baryon chiral perturbation theory (BChPT) in the hyperon decays. Finally, we discuss the importance of renormalization group evolution~(RGE) effects  in connecting  new physics models to the low energy effective field theory at  the  electroweak  scale.

In Sec.~II, we first provide predictions for the branching ratios of the neutrino modes in the SM and study their sensitivity to effective operators of the weak Hamiltonian in the presence of NP. In Sec.~III we discuss the bounds obtained from the $B_1\to B_2\nu\bar{\nu}$ decays, which
are compared with those stemming from the kaon decays $K\to \pi\nu\bar\nu$ and $K\to\pi\pi\nu\bar\nu$.
Decays into charged leptons are dominated by long-distance contributions but are also phenomenologically richer, with observables in the angular distributions potentially sensitive to short-distance structure. In Sec.~IV we show that, indeed, the leptonic forward-backward asymmetry $A_{FB}$ is such an observable~\cite{Meinel:2017ggx,He:2018yzu} and it is sensitive to the interference between the long- and short-distance contributions. Finally, we present predictions and compare the sensitivity to NP of the $\Sigma^+\to p\mu^+\mu^-$ decay to that of the rare kaon decays. In Sec.V, we discuss the importance of renormalization group evolution effects. A short summary and outlook is given in Sec.~VI.

\section{Low-energy effective Hamiltonian}
\label{sec:Heff}
Semileptonic rare $s\to d$ decays in the SM are described by the following effective Hamiltonian~\cite{Buchalla:1998ba,Cirigliano:2011ny}
\begin{align}
 \label{eq:stodLag}
 \mathcal H_{\rm eff}=\frac{G_F}{\sqrt{2}}\lambda_t\left(\sum_{i=1}^{10} C_iO_i+\sum_{\ell=e,\mu,\tau}C_{\nu_\ell}^LO_{\nu_\ell}^L\right),
\end{align}
where $G_F$ is the Fermi constant, $C_i$ and $C_{\nu_\ell}$ are the Wilson coefficients (WCs), and $\lambda_q=V_{qs}V_{qd}^*$ is defined by the product of the two Cabibbo-Kobayashi-Maskawa (CKM) matrix elements. The subscript $q$ represents $u$, $c$, or $t$ quark. The operators $O_{1,2}$ are the ``current-current'' Fermi interactions from tree-level $W$ exchanges, $O_{3-6}$ are the QCD penguin-loop contributions, and $O_8$ is the ``chromomagnetic'' operator. The remaining operators will be explained below and one can find more details in Refs.~\cite{Buchalla:1998ba,Buras:2020xsm}.

Decays into neutrinos are described by the following operator
\begin{eqnarray}
\label{eq:stodnunuOper}
 O_{\nu_\ell}^L=\alpha\left(\bar d\gamma_\mu(1-\gamma_5)s\right)\left(\bar{\nu}_\ell\gamma^\mu(1-\gamma_5)\nu_\ell\right),
\end{eqnarray}
and Wilson coefficient
\begin{eqnarray}
\label{eq:stodnunuWC}
C_{\nu_\ell}^L=\frac{1}{2\pi\sin^2\theta_w}\left(\frac{\lambda_c}{\lambda_t}X_c^\ell+X_t\right).
\end{eqnarray}
Here $\alpha$ and $\sin^2\theta_w$ denote the electromagnetic coupling and the weak mixing angle, respectively. The contributions of the charm- and top-quark are characterized by $X_c^\ell$ and $X_t$, while that of the up-quark is much suppressed by the GIM mechanism. The first term in Eq.~(\ref{eq:stodnunuWC}) has some mild sensitivity to the charged-lepton flavor in the box-diagram and therefore it has the corresponding label in the Wilson coefficient. The $\ell$ stands for $e$, $\mu$, and $\tau$ lepton respectively. In the numerical analysis, we take $X_t=1.48$ considering the next-to-leading order (NLO) QCD corrections~\cite{Buchalla:1995vs,Buchalla:1998ba} and two-loop electroweak corrections~\cite{Brod:2010hi}. For  $X_c^\ell$, we take $X_c^e\simeq X_c^\mu=1.04\times10^{-3}$ and $X_c^\tau=0.70\times10^{-3}$ obtained by the renormalization group (RG) calculation with the next-to-leading logarithmic approximation~\cite{Buchalla:1995vs,Buchalla:1998ba}.~\footnote{We note  that the combination of $\frac{1}{\lambda^4}(\frac{2}{3}X_c^e+\frac{1}{3}X_c^\tau)=0.36$ is consistent with that determined in Refs.~\cite{Buras:2005gr,Buras:2006gb,Brod:2008ss}, which  considered NNLO QCD and two-loop electroweak corrections. } The $\alpha/\sin^2\theta_w$ is a renormalization scheme dependent quantity. The scheme dependence can be removed by considering higher order electroweak effects in $K\to\pi\nu\bar{\nu}$. Here we adopt the modified minimal subtraction scheme ($\overline{\rm MS}$) definitions of $\alpha$ and $\sin^2\theta_w$, i.e., $\alpha=\alpha_{\overline{\rm MS}}(M_Z)=1/127.952$ and $\sin^2\theta_w=\sin^2\hat{\theta}_w^{\overline{\rm MS}}(M_Z)=0.23121$ from the Particle Data Group (PDG) review~\cite{ParticleDataGroup:2020ssz}. These higher order electroweak corrections in $K\to\pi\nu\bar{\nu}$ with the $\overline{\rm MS}$ scheme are found below 2\%~\cite{Buchalla:1997kz,Brod:2010hi}. We take the Wolfenstein parameters for the CKM matrix and the Fermi constant $G_F$ from the most recent pdgLive~\cite{ParticleDataGroup:2020ssz}.

The four-quark operators contributing to the $s\to d \ell^+\ell^-$ transitions, $O_{7,9,10}$, are defined as follows
\begin{eqnarray}
&&\qquad\qquad\qquad O_7=i\frac{e}{4\pi}m_s\bar{d}\sigma_{\mu\nu}(1+\gamma_5)sF^{\mu\nu},\nonumber\\
&&O_9=\alpha\left(\bar{d}\gamma_\mu(1-\gamma_5)s\right)\left(\bar{\ell}\gamma^\mu\ell\right),\qquad O_{10}=\alpha\left(\bar{d}\gamma_\mu(1-\gamma_5)s\right)\left(\bar{\ell}\gamma^\mu\gamma_5\ell\right),\label{Eq:shortstodmumu}
\end{eqnarray}
where $\alpha=e^2/4\pi$ and $\sigma_{\mu\nu}=\left[\gamma_\mu,\gamma_\nu\right]/2$~\cite{Kadeer:2005aq}. These Lagrangians stem from electromagnetic penguin-, electroweak penguin-, and box-loop diagrams. The Wilson coefficients can be expressed as $C_7=\frac{C_\gamma^c}{4\pi}\frac{\lambda_c}{\lambda_t}$, $C_9=z_{7V}\frac{\lambda_u}{\lambda_t}-y_{7V}$, and $C_{10}=-y_{7A}$. The Wilson coefficient $C_7$ is dominated by the charm penguin-loop contribution. The terms $y_{7V,7A}$ are produced by integrating out the top quark at $\mu=m_W$ while $z_{7V}(\mu)$ are the corresponding ones by integrating out the charm quark at $\mu=m_c$. Both coefficients $z_{7V}$ and  $y_{7V}$ depend on the
renormalization scale through the mixing of four-quark operators in QCD, which must be canceled at the observable level by  long-distance contributions. The axial vector coupling to the leptons, $y_{7A}$, is not renormalized in QCD. The long-distance interactions dominating the $s\to d \ell^+\ell^-$ modes couple vectorially to the leptons and are difficult to calculate from first principles. Specifically, the long-distance amplitudes for the $B_1\to B_2\ell^+\ell^-$ decays can be parameterized as~\cite{He:2005yn}
\begin{eqnarray}
{\cal M}_{\rm LD}=-\frac{e^2G_F}{q^2}\bar{B}_2\sigma_{\mu\nu}q^\nu(a+b\gamma_5)B_1\bar{\ell}^-\gamma^\mu\ell^+
-e^2G_F\bar{B}_2\gamma_\mu(c+d\gamma_5)B_1\bar{\ell}^-\gamma^\mu\ell^+,\label{laglong}
\end{eqnarray}
where $a$, $b$, $c$ and $d$ are complex form factors, and $B_1$ and $B_2$ denote the initial and final octet baryons, respectively. For the WCs in the $s\to d \ell^+\ell^-$ transitions, we use the NLO QCD values $z_{7V}(\text{1 GeV})=-0.046$, $y_{7V}(\text{1 GeV})=0.735$, and $y_{7A}=y_{7A}(m_W)=-0.700$ obtained by choosing the ``naive dimensional regularization'' scheme with $\gamma_5$ taken to be anticommuting~\cite{Buchalla:1995vs}.~\footnote{Another possible choice is the 't-Hooft-Veltman scheme with non-anticommuting $\gamma_5$.} It has been pointed out in Ref.~\cite{Buchalla:1995vs} that the operator $O_7$ does not influence the WCs of other operators since its contribution is suppressed by an additional factor $m_s$. Here we take the value $C_\gamma^c=0.13$ derived from the  short-distance contribution of the weak radiative transition $s\to d\gamma$~\cite{Shifman:1976de,He:2005yn}.

Beyond the SM and for scenarios without  right-handed neutrinos, the only new operator that can contribute to $s\to d\nu\bar\nu$ is $O_{\nu_\ell}^R$, which is the same as Eq.~\eqref{eq:stodnunuOper} but with a chiral flip in the quark current. In the decays into charged leptons, we can also have semileptonic operators with a chiral flip in the quark current ($O_{9,10}'$) or (pseudo) scalar operators, but not tensors if NP is significantly heavier than the electroweak scale~\cite{Alonso:2014csa}.~\footnote{In our convention, these operators are written as follows:
\begin{eqnarray}
&&O_S=\alpha\left(\bar{d}(1+\gamma_5)s\right)\left(\bar{\ell}\ell\right),\qquad O_S'=\alpha\left(\bar{d}(1-\gamma_5)s\right)\left(\bar{\ell}\ell\right),\nonumber\\
&&O_P=\alpha\left(\bar{d}(1+\gamma_5)s\right)\left(\bar{\ell}\gamma_5\ell\right),\qquad O_P'=\alpha\left(\bar{d}(1-\gamma_5)s\right)\left(\bar{\ell}\gamma_5\ell\right).\nonumber
\end{eqnarray}} However, tensor operators are nonvanishing in the  nonlinear effective field theory of Ref.~\cite{Cata:2015lta}. In this work, we assume that all the WCs are real as we refrain from discussing CP violation.

\section{$s\to d\nu\bar{\nu}$ processes}
\label{sec:hyps_nus}
\subsection{$B_1\to B_2\nu\bar{\nu}$ decays}
Let us first investigate the rare hyperon decays into neutrinos. As pointed out in Refs.~\cite{Hu:2018luj,Yang:2015era}, the
vector transition matrix elements for the decuplet-to-octet transitions are not well determined because of  lack of experimental data for the electric-quadrupole and magnetic-dipole  transition form factors. Therefore, in the following we  only study  five rare octet-hyperon decays: $\Lambda\to n\nu\bar{\nu}$, $\Sigma^{+}\to p\nu\bar{\nu}$, $\Xi^{-}\to\Sigma^{-}\nu\bar{\nu}$, $\Xi^{0}\to\Sigma^{0}\nu\bar{\nu}$, and $\Xi^{0}\to\Lambda\nu\bar{\nu}$. Following Ref.~\cite{Chang:2014iba}, one can obtain the total decay width in the presence of NP, which, expanded up to NLO in terms of the SU(3) breaking parameter $\delta$, reads
{\small
\begin{eqnarray}
\Gamma=\sum_{\ell=e,\mu,\tau}\frac{\alpha^2G_F^2|\lambda_t|^2f_1(0)^2\Delta^5}{60\pi^3}\cdot
\left[\left(1-\frac{3}{2}\delta\right)\left|C_{\nu_{\ell}}^L+C_{\nu_{\ell}}^R\right|^2+3\left(1-\frac{3}{2}\delta\right)\frac{g_1(0)^2}{f_1(0)^2}
\left|C_{\nu_{\ell}}^L-C_{\nu_{\ell}}^R\right|^2+{\cal O}\left(\delta^2\right)\right],\label{eq:B1toB2nunuGamma}
\end{eqnarray}
}
where $\Delta=M_1-M_2$ and $\delta=\Delta/M_1$. NP can enter with both chiralities $C_{\nu_\ell}^L=C_{\nu_\ell}^{L, \rm SM}+\delta C_{\nu_\ell}^L$ and $C_{\nu_{\ell}}^R$. The parameterization of relevant hadronic matrix elements can be found in Appendix A.

The $B_1\to B_2$ form factors can be related to those entering  the semileptonic hyperon decays $B_1\to B_2\ell^-\bar{\nu}$~\cite{Cabibbo:2003cu}  using isospin symmetry. As shown in Eq.~(\ref{eq:B1toB2nunuGamma}), only the vector and axial vector couplings, $f_1(0)$ and $g_1(0)$, are necessary for our predictions.~\footnote{The $q^2$-dependent scales can be expanded in powers of $q^2/M_X^2\sim\delta^2$, where $M_X\sim1~{\rm GeV}$ is a hadronic scale related to the mass of the resonance coupling to a given current~\cite{Ecker:1989yg}.} For consistency, the  axial vector couplings $g_1(0)$ are obtained by including the ${\cal O}(\delta)$ corrections calculated in chiral perturbation theory~\cite{Ledwig:2014rfa}. The vector couplings $f_1(0)$ are the SU(3) symmetric values which are protected from $\mathcal O(\delta)$ corrections by the Ademollo-Gatto theorem~\cite{Ademollo:1964sr}. For the sake of convenience, we collect these results in Table~\ref{tab:HyperFFs}. In addition, at $\mathcal O(\delta)$ one has in Eq.~(\ref{eq:B1toB2nunuGamma}) the contribution from the weak-magnetic form factor $g_2(0)$, which is neglected because $g_2(0)\sim\mathcal O (\delta)$.
\begin{table}[h!]
\renewcommand{\arraystretch}{1.2}
\begin{center}
    \begin{tabular}{|cccccc|}
      \hline
      &$\Lambda n$ & $\Sigma^+ p$ & $\Xi^-\Sigma^-$ & $\Xi^0\Sigma^0$ & $\Xi^0\Lambda$\\
\hline
$f_1(0)$ &$-1.22(6)$ & $-1.00(5)$ & $1.00(5)$ &$-0.71(4)$&$1.22(6)$\\
$g_1(0)$ &$-0.88(2)$&$0.33(2)$&$1.22(4)$&$-0.86(3)$&$0.21(4)$ \\
      \hline
    \end{tabular}
  \end{center}
\caption{\label{tab:HyperFFs} Values of vector and axial vector couplings, $f_1(0)$ and $g_1(0)$, for the $B_1\to B_2\nu\bar{\nu}$ decays studied in this work. The uncertainty in $g_1(0)$ stems from experimental data~\cite{Ledwig:2014rfa}. The one in $f_1(0)$ corresponds to $\mathcal O(\delta^2)$ corrections estimated as a 5\% relative uncertainty by comparing the SU(3)-symmetric prediction with explicit calculations in LQCD~\cite{Sasaki:2017jue}.}
\end{table}

In Table~\ref{tab:ResB1toB2nunu}, we show the results for the semileptonic rare hyperon decays into neutrinos. We first show the predicted branching fractions of $B_1\to B_2\nu\bar\nu$ valid up to NLO corrections in the SU(3)-flavor-breaking expansion. The theoretical uncertainties of these branching fractions originate from the naive-dimensional estimate of the missing $\mathcal O(\delta^2)$ terms in Eq.~(\ref{eq:B1toB2nunuGamma}). In addition, we find that our  branching fractions of  the $B_1\to B_2\nu\bar\nu$ decays in the SM are $10\%\sim60\%$ smaller  than those given in Refs.~\cite{Tandean:2019tkm,Su:2019tjn,Li:2019cbk,Hu:2018luj}. This could be attributed to  the improved nonperturbative inputs for the form factors.
\begin{table}[h!]
\renewcommand{\arraystretch}{1.2}
\begin{center}
    \begin{tabular}{|cccccc|}
      \hline
      Decay modes& ~~~~~$\Lambda n$~~~~~& ~~~~~$\Sigma^{+} p$~~~~& ~~~~~$\Xi^{-} \Sigma^{-}$~~~~~&~~~~~$\Xi^{0} \Sigma^{0}$~~~~~&~~~~~$\Xi^{0} \Lambda$~~~\\
      \hline
     $10^{13}\times{\rm BR}(B_1\to B_2\nu\bar{\nu})^{\rm SM}$ & $6.26(16)$ & $3.49(16)$ & $1.10(1)$ & $0.89(1)$ &  $5.52(13)$\\
      \hline
       $10^{6}\times{\rm BR}(B_1\to B_2\nu\bar{\nu})^{\rm BESIII}$ & $<0.3$ & $<0.4$ & $-$ & $<0.9$ & $<0.8$ \\
      \hline
      $10^{3}\times\left|\delta C_{\nu_\ell}^L+C_{\nu_\ell}^R\right|$ & $<1.6$ & $<1.7$ & $-$ & $<10$ & $<1.8$\\
      $10^{3}\times\left|\delta C_{\nu_\ell}^L-C_{\nu_\ell}^R\right|$ & $<1.3$ & $<3.4$ & $-$ & $<5.2$ & $<8.6$\\
      \hline
    \end{tabular}
  \end{center}
\caption{\label{tab:ResB1toB2nunu}Predicted branching fractions of the $B_1\to B_2\nu\bar\nu$ decays valid up to NLO corrections in the SU(3)-flavor-breaking expansion. The combinations of the two Wilson coefficients are constrained by the anticipated BESIII data~\cite{Li:2016tlt} at 90\% CL. The theoretical uncertainties of branching fractions come from the missing $\mathcal O(\delta^2)$ terms in Eq.~(\ref{eq:B1toB2nunuGamma}).}
\end{table}

Next, we present the ranges of NP contributions at 90\% CL  allowed by the constraints of the BESIII projections for the near future~\cite{Li:2016tlt}. One sees that the BESIII projections do not include the $\Xi^-\to\Sigma^-\nu\bar{\nu}$ decay because it cannot be detected at BESIII due to the ineffectiveness of the ``tag technique''~\cite{Li:2016tlt} for this mode. From the constraints of the anticipated BESIII data, we find that among the five $B_1\to B_2\nu\bar\nu$ decays, the $\Lambda\to n\nu\bar{\nu}$ decay  would be the one most sensitive to new physics.

\subsection{$K\to\pi\nu\bar{\nu}$ and $K\to\pi\pi\nu\bar{\nu}$ decays}

The $K\to\pi\nu\bar{\nu}$ and $K\to\pi\pi\nu\bar{\nu}$ decays also probe the $s\to d \nu\bar\nu$ processes at the quark level.
Theoretically, the form factors of $K^+\to\pi^+\nu\bar{\nu}$ and $K_L\to\pi^0\nu\bar{\nu}$  can be related via isospin symmetry to the semileptonic charged-current $K^+\to\pi^0 e^+\nu_e$ decay, which are well known experimentally. Using the effective Hamiltonian of Eq.~(\ref{eq:stodLag}) and applying the isospin symmetry relations (see, e.g., Ref.~\cite{Buras:2020xsm}), one can easily obtain the branching fractions of the two $K\to\pi\nu\bar{\nu}$ processes in the presence of NP as
\begin{eqnarray}
{\rm BR}(K^+\to\pi^+\nu\bar{\nu})=\frac{2\alpha^2|\lambda_t|^2{\rm BR}(K^+\to\pi^0 e^+\nu_e)}{|V_{us}|^2}\sum_{\ell=e,\mu,\tau}|C_{\nu_\ell}^L+C_{\nu_\ell}^R|^2,\label{eq:stodnunu3bodyKBr1}\\
{\rm BR}(K_L\to\pi^0\nu\bar{\nu})=\frac{2\alpha^2\tau_{K_L}{\rm BR}(K^+\to\pi^0 e^+\nu_e)}{\tau_{K^+}|V_{us}|^2}\sum_{\ell=e,\mu,\tau}\left({\rm Im}[(C_{\nu_\ell}^L+C_{\nu_\ell}^R)\lambda_t]\right)^2,\label{eq:stodnunu3bodyKBr2}
\end{eqnarray}
where $\tau_{K_L}$ and $\tau_{K^+}$ are the lifetimes of the $K_L$ and $K^+$ mesons and $\lambda_t=V_{ts}V_{td}^*$. For the branching fraction of $K^+\to\pi^0 e^+\nu_e$, we take the value $(5.07\pm0.04)\times10^{-2}$ from the PDG average~\cite{ParticleDataGroup:2020ssz}. Throughout the present work, we do not consider isospin breaking effects.

In this work, we study the $K^+\to\pi^+\pi^0\nu\bar{\nu}$ and $K_L\to\pi^0\pi^0\nu\bar{\nu}$ decays in  helicity basis (details can be found in Appendix B). The form factors for the $K_L\to\pi^0\pi^0$ and $K^+\to\pi^+\pi^0$ transitions only contain $I=0$ and $I=1$ components (of the $\pi\pi$ pair) for $S$- and $P$-waves, respectively. The contributions of $D$- and higher partial waves between the two pions in these form factors have been neglected. These form factors can be connected to those measured in the charged current  $K^+\to\pi^+\pi^-e^+\nu$ decay by isospin symmetry
as follows~\cite{Littenberg:1995zy,Chiang:2000bg,MartinCamalich:2020dfe}:
\begin{eqnarray}
&&\langle \pi^+\pi^0|(\bar{s}d)_{V-A}|K^+\rangle=-\sqrt{2}\langle (\pi^+\pi^-)_{I=1}|(\bar{s}u)_{V-A}|K^+\rangle,\nonumber\\
&&\langle \pi^0\pi^0|(\bar{s}d)_{V-A}|K^0\rangle=\langle (\pi^+\pi^-)_{I=0}|(\bar{s}u)_{V-A}|K^+\rangle.
\end{eqnarray}
All the form factors for the $K^+\to\pi^+\pi^-$ decay can be described as a series expansion of the dimensionless invariants $q^2=s_\pi/4m_\pi^2-1$ and $s_\ell/4m_\pi^2$, and the coefficients of relevant expansions, as well as the difference of the rescattering phases $\delta=\delta_s-\delta_p$ have been determined experimentally from the angular analysis of the $K^+\to\pi^+\pi^-e^+\nu$ decay~\cite{NA482:2010dug,NA482:2012cho}.~\footnote{See also Ref.~\cite{NA482:2014qtc} for the analysis of $K^+\to\pi^0\pi^0$, but this process has less statistics and we will not consider it in our present study.}

Therefore, the branching fractions of the four-body $K\to\pi\pi\nu\bar{\nu}$ decays as functions of the two Wilson coefficients can be expressed as
\begin{eqnarray}
&&{\rm BR}(K^+\to\pi^+\pi^0\nu\bar{\nu})=\sum_{\ell=e,\mu,\tau}\left(3.13\times10^{-3}~\left|C_{\nu_\ell}^L+C_{\nu_\ell}^R\right|^2+1.41~\left|C_{\nu_\ell}^L-C_{\nu_\ell}^R\right|^2\right)\times10^{-15},\label{BrKptopippi0}\\
&&{\rm BR}(K_L\to\pi^0\pi^0\nu\bar{\nu})=\sum_{\ell=e,\mu,\tau}\left({\rm Re}\left[\left(-6.35-2.72i\right)\left(C_{\nu_\ell}^L-C_{\nu_\ell}^R\right)\right]\right)^2\times10^{-15},\label{BrKLtopi0pi0}
\end{eqnarray}
where the numerical coefficients have been obtained by integrating over phase space. In Eqs.~(\ref{BrKptopippi0}) and (\ref{BrKLtopi0pi0}), we do not include hadronic uncertainties stemming from form factors. Compared with the exact numerical calculation using the formulae given in Appendix B, we find that the hadronic uncertainties induce effects of less than 1\%  on the branching fractions of the $K\to\pi\pi\nu\bar{\nu}$ decays.

In Table~\ref{tab:ResKtonunu}, we show the branching fractions of the $K\to\pi\nu\bar{\nu}$ and $K\to\pi\pi\nu\bar{\nu}$ decays in the SM and the ranges of NP allowed by the experiment measurements at 90\% CL. For the two $K\to\pi\nu\bar{\nu}$ decays, the SM predictions are consistent with those of Ref.~\cite{Buras:2015qea} and are close to the current experiment measurements. From Eqs.~(\ref{eq:stodnunu3bodyKBr1}) and (\ref{eq:stodnunu3bodyKBr2}), one can easily see that the theoretical uncertainties for the branching fractions of the $K\to\pi\nu\bar{\nu}$ decays originate from the current experiment measurement of ${\rm BR}(K^+\to\pi^0 e^+\nu_e)=(5.07\pm0.04)\times10^{-2}$ in the absence of isospin breaking effects. Both decay modes, especially $K^+\to\pi^+\nu\bar{\nu}$, constrain strongly $\delta C_{\nu_\ell}^L+C_{\nu_\ell}^R$ but they are not sensitive to the combination of $\delta C_{\nu_\ell}^L-C_{\nu_\ell}^R$. On the other hand, following  Ref.~\cite{Grossman:1997sk} one can easily derive by means of isospin symmetry a Grossman-Nir~(GN) bound,
\begin{eqnarray}
{\rm BR}(K_L\to\pi^0\nu\bar{\nu})_{\rm exp}\leq4.3~{\rm BR}(K^+\to\pi^+\nu\bar{\nu})_{\rm exp}.
\end{eqnarray}
Using Eq.~(\ref{Kptopipvvexp}), the standard GN bound gives ${\rm BR}(K_L\to\pi^0\nu\bar{\nu})_{\rm exp}\leq7.7\times10^{-10}$. This bound is stronger than the current direct bounds of $3.0\times10^{-9}$ and $4.9\times10^{-9}$ at 90\% CL from the KOTO experiment~\cite{KOTO:2018dsc,KOTO:2020prk}.
\begin{table}[h!]
\renewcommand{\arraystretch}{1.2}
\begin{center}
    \begin{tabular}{|ccccc|}
      \hline
      Decay modes & ~~~~~~~~$K^+ \pi^+$~~~~~~~~ & ~~~~~~~~$K_L \pi^0$~~~~~~~~& ~~~~~~~~$K^+ \pi^+\pi^0$~~~~~~~~ & ~~~~~~~~$K_L \pi^0\pi^0$~~~~~~~~\\
      \hline
      ${\rm BR}^{\rm SM}$ & $8.55(4)\times10^{-11}$ & $2.89(1)\times10^{-11}$ & $8.35(22)\times10^{-15}$ & $2.59(3)\times10^{-13}$\\

      ${\rm BR}^{\rm Expt}$  & $<1.78\times10^{-10}$ & $<3.0\times10^{-9}$ & $<4.3\times10^{-5}$ & $<8.1\times10^{-7}$\\
      \hline
      $\delta C_{\nu_\ell}^L+C_{\nu_\ell}^R$ & $(-3.4,0.6)$ & $(-11.5,9.4)$ & $(-2.2,2.2)\times10^6$ & $-$ \\

      $\delta C_{\nu_\ell}^L-C_{\nu_\ell}^R$ & $-$ & $-$ & $(-1.1,1.1)\times10^5$ & $(-2.7,2.7)\times10^3$ \\
      \hline
    \end{tabular}
  \end{center}
\caption{\label{tab:ResKtonunu}Predictions in the SM for the branching fractions of $K\to\pi\nu\bar{\nu}$ and $K\to\pi\pi\nu\bar{\nu}$. The constraints on the combinations of two Wilson coefficients are derived from the experimental measurements at 90\% CL~\cite{NA62:2020fhy,KOTO:2018dsc,E787:2000iwe,E391a:2011aa}.}
\end{table}

For the $K^+\to\pi^+\pi^0\nu\bar{\nu}$ decay, our prediction for the branching fraction in the SM is $20\%$ smaller than that of Ref.~\cite{Littenberg:1995zy} while it is $40\%$ larger than that of Ref.~\cite{Chiang:2000bg} for the $K_L\to\pi^0\pi^0\nu\bar{\nu}$ decay, due to different inputs for the form factors. The theoretical uncertainties of the $K\to\pi\pi\nu\bar{\nu}$ branching fractions in Table~\ref{tab:ResKtonunu} originate from form factors introduced in Appendix B. It is clear that the $K\to\pi\pi\nu\bar{\nu}$ decays determine poorly $\delta C_{\nu_\ell}^L+C_{\nu_\ell}^R$ but they can better constrain $\delta C_{\nu_\ell}^L-C_{\nu_\ell}^R$. In addition, the $K_L\to\pi^0\pi^0\nu\bar{\nu}$ decay provides a constraint on NP more stringent than the $K^+\to \pi^+\pi^0\nu\bar{\nu}$ decay.
\begin{figure}[h!]
  \centering
  \includegraphics[width=9cm]{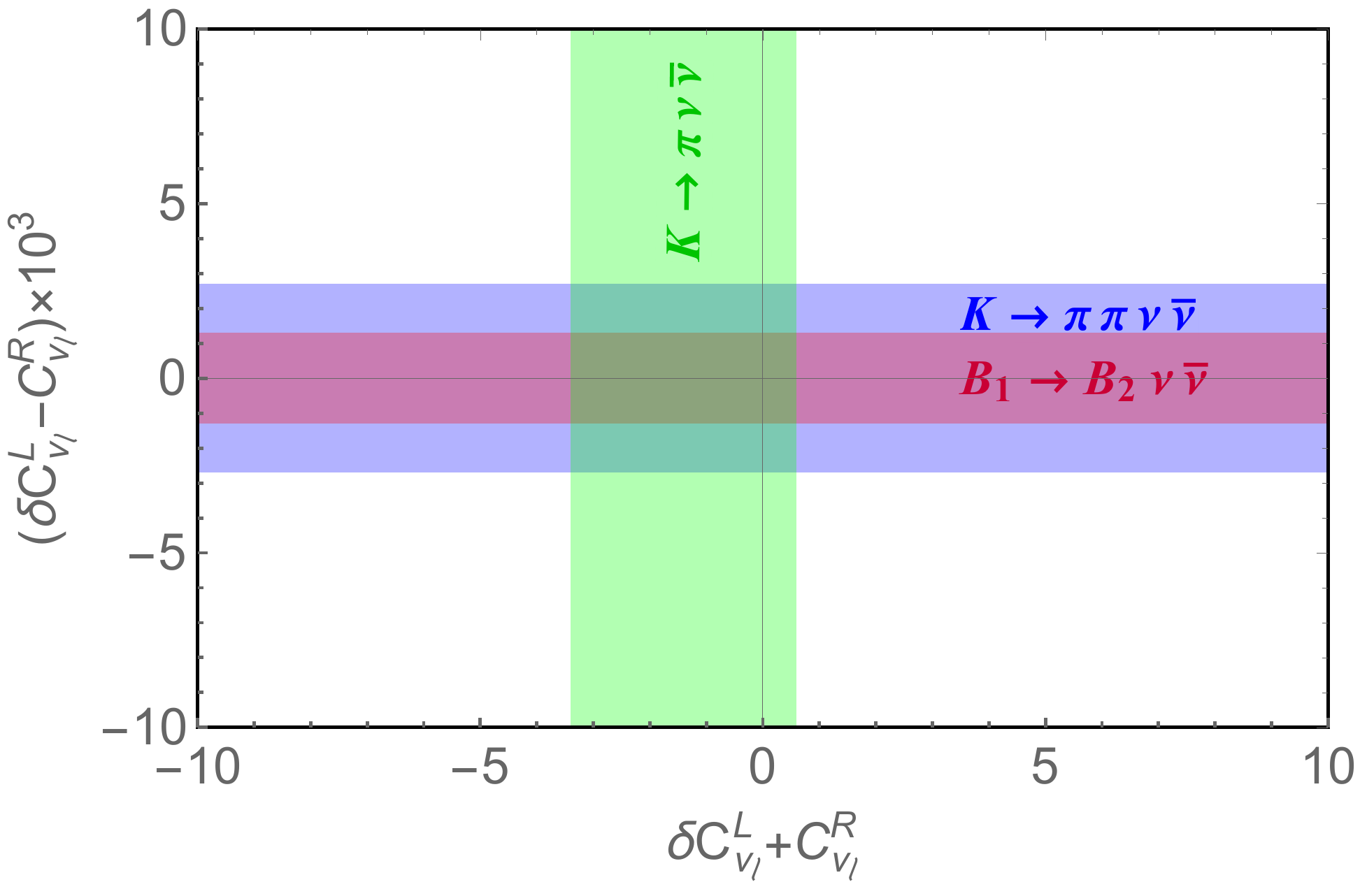}\\
  \caption{90\% CL constraints on $\delta C_{\nu_\ell}^L+C_{\nu_\ell}^R$ and $\delta C_{\nu_\ell}^L-C_{\nu_\ell}^R$ from the BESIII projections in the $B_1\to B_2\nu\bar\nu$ decays~ (region in red) and the current experiment data in the kaon decay modes~(regions in green and blue).}\label{Fig1:btosnunu}
\end{figure}

In Fig.~\ref{Fig1:btosnunu}, we compare the $B_1\to B_2\nu\bar\nu$ decays with their kaon siblings and conclude that $\delta C_{\nu_\ell}^L+C_{\nu_\ell}^R$ can be well determined by the kaon modes. However future measurements of the $B_1\to B_2\nu\bar\nu$ at BESIII could lead to better constraints  on $\delta C_{\nu_\ell}^L-C_{\nu_\ell}^R$ compared to their kaon siblings. Therefore, it would be interesting to search for these semileptonic rare hyperon decays at BESIII.

\section{$s\to d\ell^+\ell^-$ processes}
\label{sec:sdmumu}
\subsection{$B_1\to B_2\ell^+\ell^-$ decays}
In this subsection, we investigate the rare hyperon decays into charged leptons. Using Eqs.~(\ref{eq:stodLag}), (\ref{Eq:shortstodmumu}) and (\ref{laglong}) we calculate the relevant observables  in helicity basis, because then the interference between the long-distance and short-distance contributions in the leptonic forward-backward asymmetry becomes rather transparent (see Appendix C for more details). Furthermore, following Ref.~\cite{Chang:2014iba}, we expand the short-distance contributions  in terms of the SU(3)-flavor-breaking expansion up to $\mathcal O(\delta)$ and keep the long-distance amplitudes unchanged, and then we obtain
\begin{eqnarray}
\frac{d\Gamma}{d\cos\theta_\ell}={\cal N}_f\left[I_1+I_2\cos\theta_\ell+I_3\cos^2\theta_\ell\right],
\end{eqnarray}
with
\begin{eqnarray}
&&I_1=\left(\frac{137.06}{\Delta^2f_1(0)^2}\right)\left(1-\frac{3}{2}\delta\right)\left|\frac{a}{\lambda_t}\right|^2
+\left(\frac{58.50}{f_1(0)^2}\right)\left(1-\frac{3}{2}\delta\right)\left|\frac{c}{\lambda_t}\right|^2\nonumber\\
&&\qquad+\left(\frac{1221.67}{\Delta^2f_1(0)^2}\right)\left(1-\frac{3}{2}\delta\right)\left|\frac{b}{\lambda_t}\right|^2
+\left(\frac{974.60}{f_1(0)^2}\right)\left(1-\frac{3}{2}\delta\right)\left|\frac{d}{\lambda_t}\right|^2\nonumber\\
&&\qquad+\left(\frac{168.52}{\Delta f_1(0)^2}\delta\right){\rm Re}\left[\frac{ac^*}{\lambda_t\lambda_t^*}\right]
-\left(\frac{2199.79}{\Delta f_1(0)^2}\right)\left(1-\frac{3}{2}\delta\right){\rm Re}\left[\frac{bd^*}{\lambda_t\lambda_t^*}\right],\nonumber\\
&&I_2=\left(C_S+C_S'\right){\rm Re}\left[\left(\frac{2.32}{\Delta f_1(0)}\delta\right)\frac{f_S(0)}{f_1(0)}\left(\frac{a}{\lambda_t}\right)^*
+\left(\frac{4.64}{f_1(0)}\right)\left(1-\frac{3}{2}\delta\right)\frac{f_S(0)}{f_1(0)}\left(\frac{c}{\lambda_t}\right)^*\right]\nonumber\\
&&\qquad+\left(C_S-C_S'\right){\rm Re}\left[\left(\frac{2.32}{\Delta f_1(0)}\delta\right)\frac{g_P(0)}{f_1(0)}\left(\frac{b}{\lambda_t}\right)^*
-\left(\frac{2.32}{f_1(0)}\delta\right)\frac{g_P(0)}{f_1(0)}\left(\frac{d}{\lambda_t}\right)^*\right]\nonumber\\
&&\qquad+\left(C_{10}+C_{10}'\right){\rm Re}\left[\left(\frac{5.51}{\Delta f_1(0)}\delta\right)\left(\frac{f_1(0)+2f_2(0)}{f_1(0)}\right)\left(\frac{b}{\lambda_t}\right)^*
-\left(\frac{4.68}{f_1(0)}\delta\right)\left(\frac{f_1(0)+2f_2(0)}{f_1(0)}\right)\left(\frac{d}{\lambda_t}\right)^*\right]\nonumber\\
&&\qquad+\left(C_{10}-C_{10}'\right){\rm Re}\left[\left(\frac{11.02}{\Delta f_1(0)}\right)\frac{g_1(0)}{f_1(0)}\left(1-\frac{3}{2}\delta\right)\left(\frac{a}{\lambda_t}\right)^*
+\left(\frac{4.68}{f_1(0)}\delta\right)\frac{g_1(0)}{f_1(0)}\left(\frac{c}{\lambda_t}\right)^*\right],\nonumber\\
&&I_3=\left(\frac{8.30}{\Delta^2f_1(0)^2}\right)\left(1-\frac{3}{2}\delta\right)\left|\frac{a}{\lambda_t}\right|^2
-\left(\frac{6.99}{f_1(0)^2}\right)\left(1-\frac{3}{2}\delta\right)\left|\frac{c}{\lambda_t}\right|^2\nonumber\\
&&\qquad+\left(\frac{9.00}{\Delta^2f_1(0)^2}\right)\left(1-\frac{3}{2}\delta\right)\left|\frac{b}{\lambda_t}\right|^2
-\left(\frac{7.00}{f_1(0)^2}\right)\left(1-\frac{3}{2}\delta\right)\left|\frac{d}{\lambda_t}\right|^2,\label{eq:B1tpB2mumuBinned}
\end{eqnarray}
where $\Delta=M_1-M_2$, $\delta=\Delta/M_1$ and ${\cal N}_f=G_F^2\alpha^2|\lambda_t|^2\Delta^5f_1(0)^2/(2048\pi^3)$. The NP contributions are characterized by $\delta C_i=C_i-C_i^{\rm SM}$ and $C_i'$. We have neglected short-distance contributions  when long-distance contributions are present. The $\theta_\ell$ is the angle between the three-momenta of the final state particle $B_2$ and the direction of flight of the  $\ell^-$ measured in the dilepton rest frame. The local form factors $f_1$, $f_2$, $g_1$, $f_S$, and $g_P$ appear in the parameterized hadronic matrix elements  explicitly given in  Appendix A.  The form factors $a$, $b$, $c$, and $d$ characterizing long-distance contributions stem from Eq.~(\ref{laglong}). The decay width and leptonic forward-backward asymmetry  for the $B_1\to B_2\ell^+\ell^-$ decays can be expressed in terms of $I_1$, $I_2$, and $I_3$, which are
\begin{eqnarray}
&&\Gamma_{\Sigma^+}=2{\cal N}_f(I_1+\frac{1}{3}I_3),\\
&&A_{FB,\Sigma^+}=\frac{{\cal N}_f\cdot I_2}{\Gamma_{\Sigma^+}}.\label{Eq:sigmatopmumuAFB}
\end{eqnarray}

In particular, for the $\Sigma^+\to p\mu^+\mu^-$ decay, the form factors $f_S(0)$ and $g_P(0)$ from short-distance contributions can be obtained using the conservation of vector current~(CVC) and the partial conservation of axial current~(PCAC) in QCD:
\begin{eqnarray}
&&f_S(0)=\frac{M_{\Sigma^+}-M_p}{m_s-m_d}f_1(0),\\ &&g_P(0)=\frac{M_{\Sigma^+}+M_p}{m_s+m_d}g_1(0),
\end{eqnarray}
where $f_1(0)$ and $g_1(0)$ are listed in Table~\ref{tab:HyperFFs}. The $f_2(0)/f_1(0)$ has been measured at Fermilab~\cite{Hsueh:1988ar}. Nevertheless, we take the SU(3) symmetric prediction $f_2(0)=-(2\mu_n+\mu_p-1)$ in the following analysis, where $\mu_p$ and $\mu_n$ are the proton and neutron anomalous magnetic moments which are calculated precisely up to NLO in
covariant baryon chiral perturbation theory (BChPT) with the extended-on-mass-shell renormalization~(EOMS) scheme~\cite{Geng:2008mf,Geng:2009hh}. All the form factors above are evaluated at $\mu=1$~GeV.

In addition, the four complex form factors $a$, $b$, $c$, and $d$ which stem from long-distance contributions have been estimated in the heavy baryon chiral perturbation theory~\cite{Jenkins:1992ab}, relativistic baryon chiral perturbation theory~\cite{Neufeld:1992np}, and vector-meson-dominance model~\cite{He:2005yn}. In our analysis we take the latest relativistic results used in Refs.~\cite{He:2005yn,He:2018yzu}, where $c(0)=(-0.65+2.75i)\times10^{-2}$ and $d(0)=(-83+3.8i)\times10^{-4}$. The real parts of the form factors $a$ and $b$ cannot be uniquely determined at present from experimental data. For ${\rm Re}[a(q^2)]$ and ${\rm Re}[b(q^2)]$, their values at $q^2=0$ can be extracted
from the experimental branching fractions and asymmetry parameters  using their imaginary parts as inputs. Therefore, it yields four sets of solutions for $a$ and $b$, which are listed in Table~\ref{tab:ResSigmatopObs}.~\footnote{In Ref.~\cite{He:2005yn}, it is shown that the form factors $a$, $b$, $c$ and $d$ have fairly mild $q^2$-dependence. Therefore, it is reasonable to use  their values at $q^2=0$ in our analysis.}
\begin{table}[h!]
\renewcommand{\arraystretch}{1.2}
\begin{center}
    \begin{tabular}{|ccccc|}
      \hline
       & $a\times10^{-3}$ & $b\times10^{-3}$ & ${\rm BR}\times10^8$ & $A_{FB}\times10^5$\\
      \hline
      Case 1   & $13.3+2.85i$ & $-6.0-1.84i$ & $1.6$ & $-1.4$\\

      Case 2  & $-13.3+2.85i$ & $6.0-1.84i$  & $3.3$ & $0.2$\\

      Case 3 & $6.0+2.85i$ & $-13.3-1.84i$ & $5.2$ & $0.6$\\

      Case 4 & $-6.0+2.85i$ & $13.3-1.84i$ & $8.9$ & $-0.5$\\
      \hline
    \end{tabular}
  \end{center}
\caption{\label{tab:ResSigmatopObs}Predicted branching fraction and leptonic forward-backward asymmetry $A_{FB}$ in the SM for the $\Sigma^+\to p \mu^+\mu^-$ decay, where $a$ and $b$ are in units of GeV.}
\end{table}

In Table~\ref{tab:ResSigmatopObs}, we predict the branching fraction and the leptonic forward-backward asymmetry $A_{FB}$ in the SM for the $\Sigma^+\to p\mu^+\mu^-$ decay, which are
\begin{eqnarray}
&&1.6\times10^{-8}\leq{\rm BR}(\Sigma^+\to p\mu^+\mu^-)\leq8.9\times10^{-8},\\
&&-1.4\times10^{-5}\leq A_{FB}(\Sigma^+\to p\mu^+\mu^-)\leq0.6\times10^{-5},
\end{eqnarray}
where the theoretical uncertainties originate mainly from long-distance form factors $a$, $b$, $c$, and $d$. Compared with the results of Refs.~\cite{He:2005yn,He:2018yzu}, one finds that the predictions for the branching fraction are consistent, however there is a sign difference  for the leptonic forward-backward asymmetry $A_{FB}$ in cases 1 and 2 due to the different definitions for the angle $\theta_\ell$.
In addition, we note that our predicted  $A_{FB}$ are smaller than those of Refs.~\cite{He:2005yn,He:2018yzu} in cases 3 and  4. The reason can be traced back to the fact that $f_2(0)$, multiplied by the long-distance contribution, is counted as of order $\delta$ (i.e. NLO) in the present work, as can be seen in Eq.~(\ref{eq:B1tpB2mumuBinned}). However,  the contribution from $f_2(0)$ is not included in Refs.~\cite{He:2005yn,He:2018yzu}.~\footnote{If we set $f_2(0)=0$ in Eq.~(\ref{eq:B1tpB2mumuBinned}), we recover the results of Refs.~\cite{He:2005yn,He:2018yzu}.} We also find that only the branching fraction predicted in case 1 and case 2 are in agreement with the current LHCb's BR measurement~\cite{LHCb:2017rdd}. Therefore, in the following, we focus on the contributions of NP to $A_{FB}$ with the  inputs given  in cases 1 and  2.

In Ref.~\cite{He:2018yzu}, the authors have studied the sensitivity of the leptonic forward-backward asymmetry $A_{FB}$ to the two new interactions $(\bar d \gamma^\mu s)(\bar\ell\gamma_\mu\gamma_5\ell)$ and $(\bar d \gamma^\mu\gamma_5s)(\bar\ell\gamma_\mu\gamma_5\ell)$ but this was not compared with the corresponding bounds from kaon decays. From  Eq.~(\ref{eq:B1tpB2mumuBinned}), one can see that the observable $A_{FB}(\Sigma^+\to p\mu^+\mu^-)$ is sensitive to four operators $(\bar d \gamma^\mu s)(\bar\ell\gamma_\mu\gamma_5\ell)$, $(\bar d \gamma^\mu\gamma_5s)(\bar\ell\gamma_\mu\gamma_5\ell)$, $(\bar d s)(\bar\ell\ell)$ and $(\bar d\gamma_5s)(\bar\ell\ell)$ because of the interference of long-distance and short-distance contributions. In this work, we will investigate the sensitivity of $A_{FB}(\Sigma^+\to p\mu^+\mu^-)$ to the four types of new interactions and compare it with the corresponding kaon decays. For such a purpose, we study the $K_L\to\mu^+\mu^-$ and $K^+\to\pi^+\mu^+\mu^-$ decays.

\subsection{$K_L\to\mu^+\mu^-$ and $K^+\to\pi^+\mu^+\mu^-$ decays}
Following Ref.~\cite{Mescia:2006jd}, one can easily obtain the branching fraction of the $K_L\to\mu^+\mu^-$ decay as follows:
\begin{eqnarray}
{\rm BR}(K_L\to\mu^+\mu^-)&=&\left[6.7+\left(\frac{0.08\alpha\pi^2}{\sqrt{2}G_Fm_sm_\mu}{\rm Im}\left[\lambda_t^*(C_S-C_S')^*\right]\right)^2+\left(1.1(C_{10}'-C_{10})^*\right.\right.\nonumber\\
&&\left.\left.+\frac{0.10\alpha\pi^2}{\sqrt{2}G_Fm_sm_\mu}{\rm Re}\left[\lambda_t^*(C_P-C_P')^*\right]-0.2\pm0.4_{-0.5}^{+0.5}\right)^2\right]\times10^{-9}.\label{KtouuBR}
\end{eqnarray}

The differential  distribution of the $K^+\to\pi^+\mu^+\mu^-$ decay is conveniently expressed as,
\begin{eqnarray}
\frac{d\Gamma}{dq^2d\cos\theta_\mu}=\Gamma_0\beta_\mu\sqrt{\lambda(q^2)}\left[A(q^2)+B(q^2)\cos\theta_\mu+C(q^2)\cos^2\theta_\mu\right],
\end{eqnarray}
with
\begin{eqnarray}
&&A(q^2)=q^2\left(\beta_\mu^2\left|F_S(q^2)\right|^2+\left|F_P(q^2)\right|^2\right)
+\frac{\lambda(q^2)}{4}\left(\left|F_A(q^2)\right|^2+\left|F_V(q^2)\right|^2\right)\nonumber\\
&&\qquad\qquad+4m_\mu^2m_{K^+}^2\left|F_A(q^2)\right|^2+2m_\mu\left(m_{K^+}^2-m_{\pi^+}^2+q^2\right){\rm Re}\left[F_P(q^2)F_A(q^2)^*\right],\nonumber\\
&&B(q^2)=2m_\mu\beta_\mu\sqrt{\lambda(q^2)}{\rm Re}\left[F_S(q^2)F_V(q^2)^*\right],\nonumber\\
&&C(q^2)=-\frac{\beta_\mu^2\lambda(q^2)}{4}\left(\left|F_A(q^2)\right|^2+\left|F_V(q^2)\right|^2\right),
\end{eqnarray}
and
\begin{eqnarray}
&&F_V(q^2)=2\left(C_9+C_9'\right)^*f_+(q^2)+\frac{4m_s}{m_{K^+}+m_{\pi^+}}f_T(q^2)C_7^*+F_{V\gamma}(q^2),\nonumber\\
&&F_A(q^2)=2\left(C_{10}+C_{10}'\right)^*f_+(q^2),\nonumber\\
&&F_S(q^2)=\frac{m_{K^+}^2-m_{\pi^+}^2}{m_s-m_d}f_0(q^2)\left(C_S+C_S'\right)^*,\nonumber\\
&&F_P(q^2)=\frac{m_{K^+}^2-m_{\pi^+}^2}{m_s-m_d}f_0(q^2)\left(C_P+C_P'\right)^*
-2m_\mu\left(C_{10}+C_{10}'\right)^*\nonumber\\
&&\qquad\qquad\cdot\left[f_+(q^2)-\frac{m_{K^+}^2-m_{\pi^+}^2}{q^2}\left(f_0(q^2)-f_+(q^2)\right)\right],
\end{eqnarray}
where $\Gamma_0=G_F^2\alpha^2|\lambda_t|^2/(512\pi^3m_{K^+}^3)$, $\beta_\mu=\sqrt{1-4m_\mu^2/q^2}$, and $\lambda(q^2)=q^4+m_{K^+}^4+m_{\pi^+}^4-2(m_{K^+}^2m_{\pi^+}^2+m_{K^+}^2q^2+m_{\pi^+}^2q^2)$. The function $F_{V\gamma}(q^2)$ is the matrix element corresponding to the long-distance contribution, which is dominated by the one-photon exchange transition $K^+\to\pi^+\gamma^*\to\pi^+\mu^+\mu^-$. As shown in Ref.~\cite{DAmbrosio:1998gur}, one has
\begin{eqnarray}
F_{V\gamma}(q^2)=-\left[\left(a_++b_+\frac{q^2}{m_{K^+}^2}\right)
+\frac{1}{m_{K^+}^2G_F}W_+^{\pi\pi}(q^2)\right]\frac{\sqrt{2}}{2\pi\lambda_t^*},
\end{eqnarray}
and the real parameters $a_+$ and $b_+$ have been determined  by the NA48/2 experiment~\cite{NA482:2010zrc}.
The non-analytic term $W_+^{\pi\pi}(q^2)$ stands for the pion-loop contribution, and its expression can be found in Ref.~\cite{DAmbrosio:1998gur}.
The decay width  and forward-backward asymmetry are,
\begin{eqnarray}
&&\Gamma_{K^+}=2\Gamma_0\int_{q_{\rm min}^2}^{q_{\rm max}^2}\beta_\mu\sqrt{\lambda(q^2)}\left(A(q^2)+\frac{1}{3}C(q^2)\right)dq^2,\label{KtopiBR}\\
&&A_{FB,K^+}=\frac{\Gamma_0\int_{q_{\rm min}^2}^{q_{\rm max}^2}\beta_\mu\sqrt{\lambda(q^2)}B(q^2)dq^2}
{\Gamma_{K^+}},\label{KtopiAFB}
\end{eqnarray}
where $q_{\rm min}^2=4m_\mu^2$ and $q_{\rm max}^2=(m_{K^+}-m_{\pi^+})^2$.

In the above expressions, the form factors $f_+(q^2)$, $f_0(q^2)$, and $f_T(q^2)$ stem from the standard decompositions of the hadronic matrix elements~\cite{Shi:2019gxi}, and they can be related to those of the charge-current $K^+\to\pi^0$ and neutral-current $K^0\to\pi^0$ transitions using isospin symmetry~\cite{Buras:2020xsm}. We use the quadratic and linear parametrizations for the vector and scalar form-factors, respectively:
\begin{eqnarray}
&&f_+(q^2)=\sqrt{2}f_+^{K^+\pi^0}(0)\left(1+\lambda_+'\frac{q^2}{m_{\pi^+}^2}+\lambda_+''\frac{q^4}{2m_{\pi^+}^4}\right),\nonumber\\
&&f_0(q^2)=\sqrt{2}f_+^{K^+\pi^0}(0)\left(1+\lambda_0'\frac{q^2}{m_{\pi^+}^2}\right),
\end{eqnarray}
where $f_+^{K^+\pi^0}(0)=0.9706(27)$ is obtained from the $N_f=2+1+1$ lattice QCD simulations~\cite{FlavourLatticeAveragingGroup:2019iem}, and $\lambda_0'$, $\lambda_+'$, and $\lambda_+''$ can be found in the PDG review~\cite{ParticleDataGroup:2020ssz}. For the tensor form factor, we employ the result from lattice QCD~\cite{Baum:2011rm}:
\begin{eqnarray}
f_T(q^2)=\frac{2f_T^{K^0\pi^0}(0)}{1-s_T^{K^0\pi^0}q^2},
\end{eqnarray}
where $f_T^{K^0\pi^0}(0)=0.417(15)$ and $s_T^{K^0\pi^0}=1.10(14)~{\rm GeV}^{-2}$.

Using Eqs.~(\ref{KtouuBR}) and (\ref{KtopiBR}), we first obtain the branching fractions of $K_L\to\mu^+\mu^-$ and $K^+\to\pi^+\mu^+\mu^-$  in the SM,
\begin{eqnarray}
&&{\rm BR}(K_L\to\mu^+\mu^-)=7.64(73)\times10^{-9},\\
&&{\rm BR}(K^+\to\pi^+\mu^+\mu^-)=9.5(7)\times10^{-8},
\end{eqnarray}
which are consistent with the current PDG averages at $1\sigma$ confidence level~\cite{ParticleDataGroup:2020ssz},
\begin{eqnarray}
&&{\rm BR}(K_L\to\mu^+\mu^-)_{\rm exp}=6.84(11)\times10^{-9},\\
&&{\rm BR}(K^+\to\pi^+\mu^+\mu^-)_{\rm exp}=9.4(6)\times10^{-8}.
\end{eqnarray}
We need to emphasize that a precise SM prediction for the branching fraction of  the $K_L\to\mu^+\mu^-$ decay is complicated due to the presence of poorly known long-distance contributions. However, the $K_L\to\mu^+\mu^-$ decay can still be used to test the SM and probe NP in a clean way as demonstrated in Refs.~\cite{DAmbrosio:2017klp,Dery:2021mct}. In addition, from the definitions given in Eqs.~(\ref{KtopiBR}) and (\ref{KtopiAFB}), one clearly sees that the branching fraction of the $K^+\to\pi^+\mu^+\mu^-$ decay is insensitive to NP due to the pollution of long-distance hadronic effects. On the other hand, the leptonic forward-backward asymmetry $A_{FB}$ of the $K^+\to\pi^+\mu^+\mu^-$ decay provides a clean null test of the SM, becoming very sensitive to the new interaction $(\bar d s)(\bar\ell\ell)$ because of the interference of long-distance and short-distance contributions. Therefore, we study the sensitivity of the branching fraction of the $K_L\to\mu^+\mu^-$ decay and the leptonic forward-backward asymmetry $A_{FB}$ of the $K^+\to\pi^+\mu^+\mu^-$ decay to NP.
\begin{table}[h!]
\renewcommand{\arraystretch}{1.2}
 {\small
\begin{center}
    \begin{tabular}{|ccccc|}
      \hline
       & \multirow{2}{2.5cm}{${\rm BR}(K_L\to\mu^+\mu^-)$} & \multirow{2}{3cm}{$A_{FB}~(K^+\to\pi^+\mu^+\mu^-)$} & \multicolumn{2}{c|}{$A_{FB}~(\Sigma^+\to p\mu^+\mu^-)$}\\
       & & & Case 1 & Case 2\\
      \hline
      SM predictions  & $7.64(73)\times10^{-9}$ & $0$ & $-1.4\times10^{-5}$ & $0.2\times10^{-5}$\\

      Data  & $6.84(11)\times10^{-9}$ & $(-2.3,2.3)\times10^{-2}$ & \multicolumn{2}{c|}{$(-2.3,2.3)\times10^{-2}$}\\
      \hline
      $C_S+C_S'$ & $-$ & $(-1.7,1.7)$ & ~~~$(-6.8,6.8)\times10^2$~~& ~~$(-9.3,9.3)\times10^3$~~~\\

      $C_S-C_S'$ & $(-0.12,0.12)$ & $-$ & $(-1.3,1.3)\times10^3$ & $(-1.8,1.8)\times10^3$\\

      $\delta C_{10}+C_{10}'$ & $-$ & $-$ & $(-1.2,1.2)\times10^3$ & $(-1.8,1.8)\times10^3$\\

      $\delta C_{10}-C_{10}'$ & $(-2.35,0.59)$ & $-$ & $(-5.8,5.8)\times10^2$ & $(-1.5,1.5)\times10^3$\\
      \hline
    \end{tabular}
  \end{center}
  }
\caption{\label{tab:Resstodmumu}Observables predicted in the SM and constraints on NP at 90\% CL derived from the experimental measurements~\cite{NA482:2010zrc,ParticleDataGroup:2020ssz}. Here, we are assuming a hypothetical measurement of $A_{FB}~(\Sigma^+\to p\mu^+\mu^-)$ that is identical to $A_{FB}~(K^+\to\pi^+\mu^+\mu^-)$.}
\end{table}

From Table~\ref{tab:Resstodmumu} one finds that current experimental measurements in the kaon modes can constrain well different scenarios of NP.  We note that the leptonic forward-backward asymmetry $A_{FB}$ of  $K^+\to\pi^+\mu^+\mu^-$ is of the order of $\sim 10^{-2}$ if the combination of scalar WCs is $\sim{\cal O}(1)$. This conclusion is in agreement with those of Refs.~\cite{Chen:2003nz,Gao:2003wy}. However, we note that the kaon modes are not directly sensitive to all the NP structures, i.e., $\delta C_{10}+C_{10}'$. Therefore, in order to be sensitive to this combination of WCs one needs to turn to the $\Sigma^+\to p\mu^+\mu^-$ decay.
In Table~\ref{tab:Resstodmumu}  we also study the sensitivity of the leptonic forward-backward asymmetry $A_{FB}$ of the $\Sigma^+\to p\mu^+\mu^-$ decay to different NP structures and assuming a future experimental measurement equivalent to the kaonic one. We find that the sensitivity of this observable to NP can be $\sim10^{-3}$ times weaker than the bounds derived from the kaon modes.

In order to understand better the sensitivity of the hyperon decay to NP, we obtain numerical formulas for the different observables. The decay widths of $K^+\to\pi^+\mu^+\mu^-$ and $\Sigma^+\to p\mu^+\mu^-$  as functions  of $C_9$ and $C_9'$ are:
\begin{eqnarray}
&&\Gamma_{K^+}=\left(1.36\cdot10^{-6}\left|C_9+C_9'\right|^2+5.07\cdot10^{-2}\right)\times10^{-22}\text{ GeV},\\
&&\Gamma_{\Sigma^+}=\left(7.82\cdot10^{-8}\left|C_9+C_9'\right|^2+3.77\cdot10^{-8}\left|C_9-C_9'\right|^2+1.33\right)\times10^{-22}\text{ GeV},
\end{eqnarray}
where we have neglected the interference terms between long- and short-distance contributions. The long-distance contribution to the decay rate for the hyperon is roughly a factor $\sim \mathcal O(100)$ larger than for the kaon.~\footnote{The phase-space factor in the rate for $K^+\to\pi^+\mu^+\mu^-$ is $\sim10$ times larger than that for $\Sigma^+\to p\mu^+\mu^-$. Hence, the long-distance contribution to the hyperon decay is larger than that to the kaon decay by a factor $\sim$10-100 at the \textit{amplitude} level. } This must indeed be the case given that the $K^+$ has a lifetime longer than the $\Sigma^+$ by the same amount while both decays have similar BRs. On the other hand, and despite the long-distance effects it is still possible to roughly estimate the bounds on $\delta C_9+C_9'$ and $\delta C_9-C_9'$ from the branching fraction of $\Sigma^+\to p\mu^+\mu^-$. If we assume that the short-distance contribution saturates the experimental branching fraction ${\rm BR}(\Sigma^+\to p\mu^+\mu^-)_{\rm exp}=\left(2.2_{-1.3}^{+1.8}\right)\times10^{-8}$ one obtains upper limits in the ballpark of $10^2$ for $\delta C_9+C_9'$ and $\delta C_9-C_9'$.
Applying the same argument for the $K^+\to\pi^+\mu^+\mu^-$ decay one obtains an upper bound of order $10$ on $\delta C_9+C_9'$ from the experimental measurement ${\rm BR}(K^+\to\pi^+\mu^+\mu^-)_{\rm exp}=9.4(6)\times10^{-8}$.

Similarly, one can obtain numerical formulas for $A_{FB}$ of $K^+\to\pi^+\mu^+\mu^-$ and $\Sigma^+\to p\mu^+\mu^-$  as functions  of $C_S$, $C_S'$, $C_{10}$ and $C_{10}'$, which read
\begin{eqnarray}
&&A_{FB,K^+}=-1.38\times10^{-2}\left(C_S+C_S'\right),\\
&&A_{FB,\Sigma^+}=\left[3.39\left(C_S+C_S'\right)+1.76\left(C_S-C_S'\right)+1.94\left(C_{10}+C_{10}'\right)-3.96\left(C_{10}-C_{10}'\right)\right]\times10^{-5},\nonumber\\
\end{eqnarray}
where  we have chosen case 1 for the numerics of $\Sigma^+\to p\mu^+\mu^-$. As already anticipated by the results shown in Table~\ref{tab:Resstodmumu}, $A_{FB}$ for hyperons suffers from a loss of sensitivity of 2-3 orders of magnitude compared to the one of kaons. This can be partly explained by the different importance of the long-distance contributions to the amplitude, which appear linearly in the numerators while quadratically in the denominators of these \textit{normalized} observables.

\section{Renormalization group evolution effects }

As discussed in Sec.~\ref{sec:hyps_nus} and Sec.~\ref{sec:sdmumu}, hyperon decays can constrain some combinations of WCs which are not constrained by the kaon modes. It is natural to ask what  NP models could be tested by hyperon decays. For such a purpose, the ideal scenario is that NP only contributes to specific combinations of WCs at  tree level. However, it is well known that in general loop effects will change this situation. In order to investigate the importance of this we calculate the RGE effects in the standard model effective field theory~(SMEFT), which allows us to treat a wide variety of phenomena spanning different energy regimes, from Higgs physics to kaon decays, in a systematic and model-independent way. In the following analysis, we assume that the electroweak symmetry breaking is linearly realized, keeping the ${\rm SU(2)}_L\times {\rm U(1)}_Y$ relations~\cite{Alonso:2014csa}. The SMEFT Lagrangian describing the NP contributions to down-quark FCNC semi-leptonic decays is~\cite{Alonso:2015sja}
\begin{eqnarray}
{\cal L}_{\rm NP}=\frac{1}{\Lambda}\sum_i C_iQ_i,
\end{eqnarray}
with
\begin{eqnarray}
&&Q_{\ell q}^{(1),ij\alpha\beta}=\left(\bar{q}_L^j\gamma^\mu q_L^i\right)\left(\bar{\ell}_L^\beta\gamma_\mu\ell_L^\alpha\right),\qquad Q_{\ell q}^{(3),ij\alpha\beta}=\left(\bar{q}_L^j\vec{\tau}\gamma^\mu q_L^i\right)\left(\bar{\ell}_L^\beta\vec{\tau}\gamma_\mu\ell_L^\alpha\right),\nonumber\\
&&Q_{\ell d}^{ij\alpha\beta}=\left(\bar{d}_R^j\gamma^\mu d_R^i\right)\left(\bar{\ell}_L^\beta\gamma_\mu\ell_L^\alpha\right),\qquad~~~~Q_{qe}^{ij\alpha\beta}=\left(\bar{q}_L^j\gamma^\mu q_L^i\right)\left(\bar{e}_R^\beta\gamma_\mu e_R^\alpha\right),\nonumber\\
&&Q_{ed}^{ij\alpha\beta}=\left(\bar{d}_R^j\gamma^\mu d_R^i\right)\left(\bar{e}_R^\beta\gamma_\mu e_R^\alpha\right),\qquad~~~~Q_{\ell edq}^{ij\alpha\beta}=\left(\bar{\ell}_Le_R\right)\left(\bar{d}_Rq_L\right),\nonumber\\
&&Q_{\ell edq}^{\prime ij\alpha\beta}=\left(\bar{e}_R\ell_L\right)\left(\bar{q}_Ld_R\right),
\end{eqnarray}
where Greek letters $\alpha$, $\beta$ and Latin letters $i$, $j$ stand for lepton flavor indices and quark flavor indices, respectively, $q_L$ and $\ell_L$ are the quark and lepton doublets, $e_R$ and $d_R$ are the right-handed charged leptons and down-type quarks singlets. The scale $\Lambda$ is the new physics scale. Following the convention of Ref.~\cite{Buras:2020xsm}, the SM fields for quarks and leptons are denoted as $q_L\sim({\bf3},{\bf2},\frac{1}{6})$, $d_R\sim({\bf3},{\bf1},-\frac{1}{3})$, $\ell_L\sim({\bf1},{\bf2},-\frac{1}{2})$, and $e_R\sim({\bf1},{\bf1},-1)$, respectively.

For the $s\to d\mu^+\mu^-$ and $s\to d\nu\bar{\nu}$ transitions, we work in the basis where the down-type quark mass matrix is diagonal. At the electroweak scale $\upsilon=(\sqrt{2}G_F)^{-1/2}\approx246~{\rm GeV}$, the NP contributions to the Wilson coefficients in the low-energy effective Lagrangian can be expressed in terms of those in the SMEFT Lagrangian as follows:
\begin{eqnarray}
&&\left[\delta C_9\right]_{sd\mu\mu}=\frac{2\pi}{e^2\lambda_{t}}\frac{\upsilon^2}{\Lambda^2}\left[C_{\ell q}^{(1)}+C_{\ell q}^{(3)}+C_{qe}\right]_{sd\mu\mu},\qquad\left[C_9'\right]_{sd\mu\mu}=\frac{2\pi}{e^2\lambda_{t}}\frac{\upsilon^2}{\Lambda^2}\left[C_{ed}+C_{\ell d}\right]_{sd\mu\mu},\nonumber\\
&&\left[\delta C_{10}\right]_{sd\mu\mu}=\frac{2\pi}{e^2\lambda_{t}}\frac{\upsilon^2}{\Lambda^2}\left[-C_{\ell q}^{(1)}-C_{\ell q}^{(3)}+C_{qe}\right]_{sd\mu\mu},\qquad\left[C_{10}'\right]_{sd\mu\mu}=\frac{2\pi}{e^2\lambda_{t}}\frac{\upsilon^2}{\Lambda^2}\left[C_{ed}-C_{\ell d}\right]_{sd\mu\mu},\nonumber\\
&&\left[C_S\right]_{sd\mu\mu}=-\left[C_P\right]_{sd\mu\mu}=\frac{2\pi}{e^2\lambda_{t}}\frac{\upsilon^2}{\Lambda^2}\left[C_{ledq}\right]_{sd\mu\mu},\qquad~~ \left[C_S'\right]_{sd\mu\mu}=\left[C_P'\right]_{sd\mu\mu}=\frac{2\pi}{e^2\lambda_{t}}\frac{\upsilon^2}{\Lambda^2}\left[C_{ledq}'\right]_{sd\mu\mu},\nonumber\\
&&\left[\delta C_{\nu_\ell}^L\right]_{sd\nu\bar{\nu}}=\frac{2\pi}{e^2\lambda_{t}}\frac{\upsilon^2}{\Lambda^2}\left[C_{\ell q}^{(1)}-C_{\ell q}^{(3)}\right]_{sd\nu\bar{\nu}},\qquad\qquad~~~~\left[C_{\nu_\ell}^R\right]_{sd\nu\bar{\nu}}=\frac{2\pi}{e^2\lambda_{t}}\frac{\upsilon^2}{\Lambda^2}\left[C_{\ell d}\right]_{sd\nu\bar{\nu}}.
\end{eqnarray}

Let us first take the $s\to d\nu\bar{\nu}$ process as an example. We consider a $Z^\prime$ model in which a single $Z^\prime\sim({\bf1},{\bf1},0)$ gauge boson couples to left-handed leptons. The Lagrangian for this model is
\begin{eqnarray}
{\cal L}_{Z^\prime}=\left(g_L^{ij}\bar{q}_L^j\gamma^\mu q_L^i+g_R^{ij}\bar{d}_R^j\gamma^\mu d_R^i+g_L^{\alpha\beta}\bar{\ell}_L^\beta\gamma^\mu \ell_L^\alpha\right)Z^\prime_\mu,
\end{eqnarray}
which does not contribute to the WCs $C_{\ell q}^{(3)}$, $C_{qe}$, $C_{ed}$ and $C_{\ell edq}$ of the SMEFT. Furthermore, by fine-tuning the couplings $g_L^{sd}g_L^{\nu\bar{\nu}}=-g_R^{sd}g_L^{\nu\bar{\nu}}$, it yields a  purely axial coupling to quarks at tree level, i.e, $\left[\delta C_{\nu_\ell}^L(\Lambda)+C_{\nu_\ell}^R(\Lambda)\right]_{sd\nu\bar{\nu}}=0$ but $\left[\delta C_{\nu_\ell}^L(\Lambda)-C_{\nu_\ell}^R(\Lambda)\right]_{sd\nu\bar{\nu}}\neq0$. When running the RG equations given in Appendix D from the NP scale $\Lambda$ to the electroweak scale $\upsilon$ and assuming that $\Lambda=10~\upsilon$, one obtains
\begin{eqnarray}
&&\left[\delta C_{\nu_\ell}^L(\upsilon)+C_{\nu_\ell}^R(\upsilon)\right]_{sd\nu\bar{\nu}}=\frac{2\pi}{e^2\lambda_{t}}\frac{\upsilon^2}{\Lambda^2}\left[0.02C_{\ell q}^{(1)}(\Lambda)\right]_{sd\nu\bar{\nu}},\nonumber\\
&&\left[\delta C_{\nu_\ell}^L(\upsilon)-C_{\nu_\ell}^R(\upsilon)\right]_{sd\nu\bar{\nu}}=\frac{2\pi}{e^2\lambda_{t}}\frac{\upsilon^2}{\Lambda^2}\left[2.02C_{\ell q}^{(1)}(\Lambda)\right]_{sd\nu\bar{\nu}}.\label{ZmodelRGE}
\end{eqnarray}
In Eq.~(\ref{ZmodelRGE}), we see that the loop effects of the RGE generate a non-vanishing vectorial contribution at the scale $\upsilon$, which is about 1\% of that of the axial-vectorial contribution. As a result, from the  $K\to\pi\nu\bar{\nu}$ data, we obtain  a constraint  on $\delta C_{\nu_\ell}^L-C_{\nu_\ell}^R$ at the order of $10^2$, which is stronger than the direct bound of $10^3$ that could be potentially obtained by BESIII from the hyperon modes shown in Table~\ref{tab:ResB1toB2nunu}.

Next, we  investigate the $s\to d\nu\bar{\nu}$ and $s\to d \mu^+\mu^-$ processes. For this, we consider a vector leptoquark~(LQ) $U_1\sim({\bf2},{\bf1},\frac{2}{3})$. In this model, we have the following relations at the NP scale $\Lambda$
\begin{eqnarray}
\left[C_{\ell q}^{(1)}(\Lambda)\right]_{ij\alpha\beta}=\left[C_{\ell q}^{(3)}(\Lambda)\right]_{ij\alpha\beta},\qquad \left[C_{\ell d}(\Lambda)\right]_{ij\alpha\beta}=0,\qquad \left[C_{qe}(\Lambda)\right]_{ij\alpha\beta}=0.
\end{eqnarray}
It means that $U_1$ contributes exclusively to charged-lepton modes but not to neutrino modes at tree level. However, the RGE mixes both types of leptons and as a result $U_1$ will also contribute to the neutrino modes at the electroweak scale. The RGE results for the relevant combinations of WCs assuming that $\Lambda=10~\upsilon$ are as follows:
\begin{eqnarray}
&&\left[\delta C_{\nu_\ell}^L(\upsilon)+C_{\nu_\ell}^R(\upsilon)\right]_{sd\mu\mu}=\frac{2\pi}{e^2\lambda_{t}}\frac{\upsilon^2}{\Lambda^2}\left[-0.07C_{\ell q}^{(1)}(\Lambda)-0.001C_{ed}(\Lambda)\right]_{sd\mu\mu},\nonumber\\
&&\left[\delta C_{\nu_\ell}^L(\upsilon)-C_{\nu_\ell}^R(\upsilon)\right]_{sd\mu\mu}=\frac{2\pi}{e^2\lambda_{t}}\frac{\upsilon^2}{\Lambda^2}\left[-0.07C_{\ell q}^{(1)}(\Lambda)+0.001C_{ed}(\Lambda)\right]_{sd\mu\mu},\nonumber\\
&&\left[\delta C_{10}(\upsilon)+C_{10}^\prime(\upsilon)\right]_{sd\mu\mu}=\frac{2\pi}{e^2\lambda_{t}}\frac{\upsilon^2}{\Lambda^2}\left[-1.96C_{\ell q}^{(1)}(\Lambda)+0.99C_{ed}(\Lambda)\right]_{sd\mu\mu},\nonumber\\
&&\left[\delta C_{10}(\upsilon)-C_{10}^\prime(\upsilon)\right]_{sd\mu\mu}=\frac{2\pi}{e^2\lambda_{t}}\frac{\upsilon^2}{\Lambda^2}\left[-1.96C_{\ell q}^{(1)}(\Lambda)-0.99C_{ed}(\Lambda)\right]_{sd\mu\mu}.\label{U1modelRGE}
\end{eqnarray}
From Eq.~(\ref{U1modelRGE}), one can see that the loop effects of RGE from the NP scale $\Lambda$ to the eletroweak scale $\upsilon$ is at the percent level. As a result, we obtain an indirect bound on the combination of WCs $\delta C_{10}$ and $C_{10}^\prime$  at the order of $\sim10^2$ from the $K\to\pi\nu\bar{\nu}$ data. The bound is weaker than that from the $s\to d\mu^+\mu^+$ kaon modes but stronger than the anticipated bound shown in Table~\ref{tab:Resstodmumu} from the hyperon modes.

Therefore,  we conclude that loop effects from renormalization group evolution are important when connecting the low-energy EFT to new physics models. Although at present the strongest constraint on new physics come from the kaon modes, rare hyperon decays could provide  better constraints on $\delta C_{\nu_\ell}^L-C_{\nu_\ell}^R$ than the kaon modes if the upper limits from future hyperon factories can reach a sensitivity to BRs of $10^{-8}$ or lower, e.g., for the $Z^\prime$ model considered above.

\section{Conclusions and outlook}
In this work, we first predicted the branching fractions of $B_1\to B_2\nu\bar{\nu}$ in the SM using model independent inputs. More specifically, form factors, up to next-to-leading order (NLO) in the SU(3)-breaking expansion, can be obtained in covariant baryon chiral perturbation theory. We discussed the constraints on $\delta C_{\nu_\ell}^L+C_{\nu_\ell}^R$ and $\delta C_{\nu_\ell}^L-C_{\nu_\ell}^R$ derived from
the anticipated BESIII measurements in the near future, and studied the potential of the $B_1\to B_2\nu\bar{\nu}$ decays as probes of NP, in comparison with the kaon modes. We conclude that $\delta C_{\nu_\ell}^L+C_{\nu_\ell}^R$ can be well determined by the kaon mode but the $B_1\to B_2\nu\bar\nu$ decays from the BESIII measurements in the near future could be better than their kaon siblings in constraining $\delta C_{\nu_\ell}^L-C_{\nu_\ell}^R$.

In addition, the $s\to d\mu^+\mu^-$ processes, which are dominated by long-distance contributions, were investigated. We provided predictions for the branching fraction of $K_L\to\mu^+\mu^-$ and leptonic forward-backward asymmetry $A_{FB}$ of $K^+\to\pi^+\mu^+\mu^-$ and $\Sigma^+\to p\mu^+\mu^-$ in the SM. Assuming that the experimental measurements of the leptonic forward-backward asymmetry $A_{FB}$ of $K^+\to\pi^+\mu^+\mu^-$ and $\Sigma^+\to p\mu^+\mu^-$ are of the same precision, we studied the sensitivity of the $\Sigma^+\to p\mu^+\mu^-$ decay and corresponding kaon modes to different NP structures. We conclude that the current kaon bounds are a few orders of magnitude better than those of $\Sigma^+\to p\mu^+\mu^-$  except for the $\delta C_{10}+C_{10}'$ and $\delta C_9-C_9'$ scenarios.

Finally, from an analysis of the renormalization group evolution, we showed that at present the strongest constraint on new physics parameters still come from the kaon modes. From the perspective of a UV theory, it is important to consider the loop effects from renormalization group evolution when connecting the low-energy EFT to new physics models. However, hyperons could still lead to the strongest constraints on some combinations of Wilson coefficients if a sensitivity of $\sim10^{-8}$ for the branching fractions is achieved by hyperon factories in the future (e.g. at the planned super tau-charm factory~\cite{Luo:2019xqt,Peng:2020orp}).

In the present study,  the long-distance form factors $a$ and $b$ from weak radiative hyperon decays were still poorly estimated, leading to  large uncertainties to the predictions for the observables of the $\Sigma^+\to p\mu^+\mu^-$ decay. Therefore, further studies of weak radiative hyperon decays will be needed.

\acknowledgments

This work is supported in part by the National Natural Science Foundation of China under Grants No.11735003, No.11975041, and No.11961141004.
JMC acknowledges  support of the ``Flavor in the era of the LHC'' project, grant PGC2018-102016-A-I00 and of the  ``Ram\'on  y  Cajal''
program RYC-2016-20672 funded by the Spanish MINECO. Rui-Xiang  Shi acknowledges support from the National Natural Science Foundation of China under Grants No.12147145 and Project funded by China Postdoctoral Science Foundation No.2021M700343.

\appendix
\section{Form factors for the octet-to-octet transitions}
The hadronic matrix elements for the octet-to-octet transitions are parameterized in terms of form factors as follows~\cite{Chang:2014iba,Bhattacharya:2011qm}
\begin{eqnarray}
\langle{B_2 (p_2) }| \bar{d} \gamma_\mu s|{B_1 (p_1)}\rangle
&&=
\bar{u}_2 (p_2)  \left[
f_1(q^2)  \,  \gamma_\mu
+ \frac{f_2(q^2)}{M_1}   \, \sigma_{\mu \nu}   q^\nu
+ \frac{f_3(q^2)}{M_1}   \,  q_\mu
\right]
 u_1 (p_1), \nonumber\\
\langle{B_2 (p_2) }| \bar{d} \gamma_\mu \gamma_5  s |{B_1 (p_1)}\rangle
&&=
\bar{u}_2 (p_2)  \left[
g_1(q^2)    \gamma_\mu
\hspace{-0.1cm}
+ \frac{g_{2} (q^2)}{M_1}   \sigma_{\mu \nu}   q^\nu
+
\hspace{-0.1cm}
\frac{g_{3} (q^2)}{M_1}   q_\mu
\right]  \gamma_5  u_1 (p_1),  \nonumber\\
\langle{B_2 (p_2) }| \bar{d} \,   s|{B_1 (p_1)}\rangle
&&=
f_S(q^2)  \ \bar{u}_2 (p_2)  \, u_1 (p_1),  \nonumber\\
\langle{B_2 (p_2) }| \bar{d} \,  \gamma_5 \,  s| {B_1 (p_1)}\rangle
&&=
g_P(q^2)  \ \bar{u}_2 (p_2)  \, \gamma_5 \, u_1 (p_1),  \nonumber\\
\langle{B_2 (p_2) }| \bar{d} \, \sigma_{\mu \nu}  \,  s |{B_1 (p_1)}\rangle
&&=
 \bar{u}_2 (p_2)
 \left[
g_T(q^2) \, \sigma_{\mu \nu}   +  \frac{g_{T}^{(1)} (q^2)}{M_1}  \left(q_\mu \gamma_\nu - q_\nu \gamma_\mu \right)
\nonumber  \right. \\
+&&
\left.  \frac{g_{T}^{(2)} (q^2)}{M_1^2}  \left( q_\mu P_\nu - q_\nu P_\mu  \right)
+
\frac{g_{T}^{(3)} (q^2)}{M_1}
 \left(
\gamma_\mu  \slashed{q}  \gamma_\nu -
\gamma_\nu  \slashed{q} \gamma_\mu
 \right)
\right]
u_1 (p_1),\nonumber\\
\langle{B_2 (p_2) }| \bar{d} \, \sigma_{\mu \nu}\gamma_5  \,  s|{B_1 (p_1)}\rangle
&&=
 \bar{u}_2 (p_2)
 \left[
g_{T5}(q^2) \, \sigma_{\mu \nu}   +  \frac{g_{T5}^{(1)} (q^2)}{M_1}  \left(q_\mu \gamma_\nu - q_\nu \gamma_\mu \right)
\nonumber  \right. \\
+&&
\left.  \frac{g_{T5}^{(2)} (q^2)}{M_1^2}  \left( q_\mu P_\nu - q_\nu P_\mu  \right)
+
\frac{g_{T5}^{(3)} (q^2)}{M_1}
 \left(
\gamma_\mu  \slashed{q}  \gamma_\nu -
\gamma_\nu  \slashed{q} \gamma_\mu
 \right)\gamma_5
\right]
u_1 (p_1).\nonumber\\
\end{eqnarray}

\section{ $K\to\pi\pi\nu\bar{\nu}$ decays in helicity basis}
The differential decay rate for the four-body $K\to\pi\pi\nu\bar{\nu}$ decay as a function of five Cabibbo-Maksymowicz variables $s_\pi$, $s_\ell$, $\theta_\pi$, $\theta_\ell$ and $\phi$~\cite{Cabibbo:1965zzb} has the following form
\begin{eqnarray}
\frac{d^5\Gamma}{ds_\pi ds_\ell d(\cos\theta_\pi)d(\cos\theta_\ell)d\phi}=\alpha^2G_F^2{\cal N}(s_\pi,s_\ell)J(s_\pi,s_\ell,\theta_\pi,\theta_\ell,\phi),
\end{eqnarray}
where ${\cal N}(s_\pi,s_\ell)=\sigma_\pi X/(2^{15}\pi^6m_K^3)$, with $\sigma_\pi=\sqrt{1-\frac{4m_\pi^2}{s_\pi}}$ and   $X=\frac{1}{2}\lambda^{1/2}(m_K^2,s_\pi,s_\ell)$,
and  $s_\pi=(p_\pi+p_\pi)^2$, $s_\ell=(p_{\bar{\nu}}+p_\nu)^2$ and $\lambda(a,b,c)=a^2+b^2+c^2-2(ab+ac+bc)$. The ranges of five Cabibbo-Maksymowicz variables are,
\begin{eqnarray}
&&s_\pi:\qquad 4m_\pi^2\leq s_\pi\leq m_K^2,\nonumber\\
&&s_\ell:\qquad 0\leq s_\ell\leq(m_K-\sqrt{s_\pi})^2,\nonumber\\
&&\theta_\pi:\qquad 0\leq \theta_\pi\leq\pi,\nonumber\\
&&\theta_\ell:\qquad 0\leq \theta_\ell\leq\pi,\nonumber\\
&&\phi:\qquad 0\leq \phi\leq2\pi,
\end{eqnarray}
and a useful form of $J$ dependence on $\theta_\ell$ and $\phi$ is~\cite{Pais:1968zza}
\begin{eqnarray}
J&=&J_1+J_2\cos2\theta_\ell+J_3\sin^2\theta_\ell\cos2\phi+J_4\sin2\theta_\ell\cos\phi+J_5\sin\theta_\ell\cos\phi\nonumber\\
&&+J_6\cos\theta_\ell+J_7\sin\theta_\ell\sin\phi+J_8\sin2\theta_\ell\sin\phi+J_9\sin^2\theta_\ell\sin2\phi,
\end{eqnarray}
where the angular coefficients $J_i$ are functions of $s_\pi$, $s_\ell$ and $\theta_\pi$. In particular, after integrating out the angles $\theta_\ell$ and $\phi$, one notes that only $J_1$ and $J_2$ contribute to the total decay width, which is,
\begin{eqnarray}
\Gamma=\alpha^2G_F^2{\cal N}(s_\pi,s_\ell)\cdot4\pi(J_1-\frac{1}{3}J_2)\cdot \int_{4m_\pi^2}^{m_K^2}ds_\pi \int_{0}^{(m_K-\sqrt{s_\pi})^2}ds_\ell \int_{-1}^{1}d(\cos\theta_\pi),
\end{eqnarray}
where $J_1$ and $J_2$ can be explicitly  expressed in terms of the helicity amplitudes as
\begin{eqnarray}
&&J_1=\frac{s_\ell}{2}\left(3|H_+^V|^2+3|H_-^V|^2+2|H_0^V|^2+3|H_+^A|^2+3|H_-^A|^2+2|H_0^A|^2\right),\nonumber\\
&&J_2=\frac{s_\ell}{2}\left(|H_+^V|^2+|H_-^V|^2-2|H_0^V|^2+|H_+^A|^2+|H_-^A|^2-2|H_0^A|^2\right).
\end{eqnarray}

The explicit forms of these helicity amplitudes for the $K^+\to\pi^+\pi^0\nu\bar{\nu}$ decay are
\begin{eqnarray}
&&H_0^{V(A)}=\frac{i\lambda_t^*(C_{\nu_\ell}^L-C_{\nu_\ell}^R)^*\left(4F\cdot m_K^2 X+G\cdot\sigma_\pi  \cos \theta _{\pi } \left(-\left(s_l-s_{\pi }\right){}^2+m_K^4+4 X^2\right)\right)}{4 m_K^3 \sqrt{s_l}},\nonumber\\
&&H_+^{V(A)}=-\frac{i\lambda_t^*\sqrt{s_{\pi }} \sigma_\pi  \sin \theta _{\pi } \left(G\cdot m_K^2 (C_{\nu_\ell}^L-C_{\nu_\ell}^R)^*+H\cdot X (C_{\nu_\ell}^L+C_{\nu_\ell}^R)^*\right)}{\sqrt{2} m_K^3},\nonumber\\
&&H_-^{V(A)}=\frac{i\lambda_t^*\sqrt{s_{\pi }} \sigma_\pi  \sin \theta _{\pi } \left(G\cdot m_K^2 (C_{\nu_\ell}^L-C_{\nu_\ell}^R)^*-H\cdot X (C_{\nu_\ell}^L+C_{\nu_\ell}^R)^*\right)}{\sqrt{2} m_K^3},
\end{eqnarray}
and for the $K_L\to\pi^0\pi^0\nu\bar{\nu}$ decay, they are
\begin{eqnarray}
&&H_0^{V(A)}=\frac{i\sqrt{2}X\cdot F\cdot{\rm Re}\left[(C_{\nu_\ell}^L-C_{\nu_\ell}^R)\lambda_t\right]}{m_K \sqrt{s_l}},\nonumber\\
&&H_+^{V(A)}=0,\nonumber\\
&&H_-^{V(A)}=0,
\end{eqnarray}
where the complex form factors $F$, $G$ and $H$ stem from the following hadronic matrix elements~\cite{NA482:2010dug}
\begin{eqnarray}
&&\langle \pi(p_1)\pi(p_2)|\bar{s}\gamma_\mu d|K(p)\rangle=-\frac{H}{m_{K}^3}\epsilon_{\mu\nu\rho\sigma}(p_{\bar{\ell}}+p_\ell)^\nu(p_1+p_2)^\rho(p_1-p_2)^\sigma,\nonumber\\
&&\langle \pi(p_1)\pi(p_2)|\bar{s}\gamma_\mu\gamma_5 d|K(p)\rangle=-\frac{i}{m_{K}}[F(p_1+p_2)_\mu+G(p_1-p_2)_\mu+R(p_{\bar{\ell}}+p_\ell)_\mu],
\end{eqnarray}
and can be expanded in partial waves in terms of the angle $\cos\theta_\pi$~\cite{NA482:2010dug},
\begin{eqnarray}
&&F=F_se^{i\delta_s}+F_pe^{i\delta_p}\cos\theta_\pi,\nonumber\\
&&G=G_pe^{i\delta_p},\nonumber\\
&&H=H_pe^{i\delta_p}.
\end{eqnarray}

In the present work, we take $\epsilon_{0123}=-1$. The form factor $R$ enters the rate proportional to the lepton masses squared and is vanishing for our purpose.
\section{$B_1\to B_2\ell^+\ell^-$ decays in  helicity basis}
Neglecting the electromagnetic corrections, the $B_1\to B_2\ell^+\ell^-$ decay amplitude is factorized as
\begin{eqnarray}
{\cal M}=\frac{G_F\alpha}{\sqrt{2}}\lambda_t\left[H_\mu^V\langle\ell^+\ell^-|\bar{\ell}^-\gamma^\mu\ell^+|0\rangle
+H_\mu^A\langle\ell^+\ell^-|\bar{\ell}^-\gamma^\mu\gamma_5\ell^+|0\rangle+H^S\langle\ell^+\ell^-|\bar{\ell}^-\ell^+|0\rangle
+H^P\langle\ell^+\ell^-|\bar{\ell}^-\gamma_5\ell^+|0\rangle\right],\nonumber\\\label{Eq:SigmatopAmp}
\end{eqnarray}
with
\begin{eqnarray}
&&H_\mu^V=\frac{m_s}{2\pi q^2}C_7\left(\langle B_2|\bar{d}\sigma_{\mu\nu}q^\nu s|B_1\rangle+\langle B_2|\bar{d}\sigma_{\mu\nu}q^\nu\gamma_5s|B_1\rangle\right)+(C_9+C_9')\langle B_2|\bar{d}\gamma_\mu s|B_1\rangle\nonumber\\
&&\qquad~~~-(C_9-C_9')\langle B_2|\bar{d}\gamma_\mu\gamma_5s|B_1\rangle+h_\lambda(q^2),\nonumber\\
&&H_\mu^A=(C_{10}+C_{10}')\langle B_2|\bar{d}\gamma_\mu s|B_1\rangle-(C_{10}-C_{10}')\langle B_2|\bar{d}\gamma_\mu\gamma_5s|B_1\rangle,\nonumber\\
&&H^S=(C_S+C_S')\langle B_2|\bar{d}s|B_1\rangle+(C_S-C_S')\langle B_2|\bar{d}\gamma_5s|B_1\rangle,\nonumber\\
&&H^P=(C_P+C_P')\langle B_2|\bar{d}s|B_1\rangle+(C_P-C_P')\langle B_2|\bar{d}\gamma_5s|B_1\rangle,\nonumber\\
&&h_\lambda(q^2)=-\frac{4\sqrt{2}\pi}{\lambda_t}\left[\frac{1}{q^2}\bar{u}_2(k)(a+b\gamma_5)\sigma_{\mu\nu}q^\nu u_1(p)
+\bar{u}_2(k)\gamma_\mu(c+d\gamma_5) u_1(p)\right],
\end{eqnarray}

The contributions to the amplitude (\ref{Eq:SigmatopAmp}) can be projected into different angular-momentum states of the dilepton pair, characterized by its polarization vectors $\epsilon^\mu(\lambda)$ and using the completeness relation $g^{\mu\nu}=\sum g^{mn}\epsilon^\mu(m)\epsilon^{\nu*}(n)$. Here, $\lambda$ is the polarization state of the virtual gauge boson. The projections of $H_\mu^V$ and $H_\mu^A$ define the helicity amplitudes,
\begin{eqnarray}
&&H_{\lambda_2,\lambda}^V=H_\mu^V(\lambda_2)\epsilon^{\mu*}(\lambda),\nonumber\\
&&H_{\lambda_2,\lambda}^A=H_\mu^A(\lambda_2)\epsilon^{\mu*}(\lambda).
\end{eqnarray}

We do not explicitly denote the helicity of the parent hyperon $\lambda_1$ in the helicity amplitudes since
$\lambda_1$ is fixed by the relation $\lambda_1=\lambda_2-\lambda$. $H^S$ and $H^P$ contributes only to the $\lambda=t$ component.

Therefore, the amplitude in helicity basis can be written as
\begin{align}
{\cal M}=&\frac{G_F\alpha}{\sqrt{2}}\lambda_t\left\{\left[(H_{\frac{1}{2},t}^V+H_{-\frac{1}{2},t}^V)\cdot L_t^V-H_{\frac{1}{2},+}^V\cdot L_+^V-H_{-\frac{1}{2},-}^V\cdot L_-^V-(H_{\frac{1}{2},0}^V+H_{-\frac{1}{2},0}^V)\cdot L_0^V\right]\right.\nonumber\\
&+\left[(H_{\frac{1}{2},t}^A+H_{-\frac{1}{2},t}^A)\cdot L_t^A-H_{\frac{1}{2},+}^A\cdot L_+^A-H_{-\frac{1}{2},-}^A\cdot L_-^A-(H_{\frac{1}{2},0}^A+H_{-\frac{1}{2},0}^A)\cdot L_0^A\right]\nonumber\\
&+\left.(H_{\frac{1}{2},t}^S+H_{-\frac{1}{2},t}^S)\cdot L^S+(H_{\frac{1}{2},t}^P+H_{-\frac{1}{2},t}^P)\cdot L^P\right\},\label{Eq:SigmatophelAmp}
\end{align}
where
\begin{align}
g_{mn}=\left\{
\begin{array}{cc}
-1, & m=n=\pm,0,\\
1, & m=n=t,\\
0, & \rm otherwise.
\end{array}
\right.
\end{align}

Substituting the parameterized hadronic matrix elements in Appendix~A into Eq.~(\ref{Eq:SigmatophelAmp}), we obtain the decay rate for $B_1(p_1)\to B_2(p_2)\ell^+\ell^-$ as a function of $q^2$ and $\cos\theta_\ell$:
\begin{eqnarray}
\frac{d\Gamma}{dq^2d\cos\theta_\ell}={\cal N}_f(q^2)\left[I_1(q^2)+I_2(q^2)\cos\theta_\ell+I_3(q^2)\cos^2\theta_\ell\right],\label{eq:B1tpB2mumuUnbinned}
\end{eqnarray}
with
\begin{eqnarray}
I_1(q^2)=&&\left(|H_{\frac{1}{2},t}^A|^2+|H_{-\frac{1}{2},t}^A|^2\right)\cdot8m_\ell^2
+\left(|H_{\frac{1}{2},t}^S|^2+|H_{-\frac{1}{2},t}^S|^2\right)\cdot\left(2q^2-5m_\ell^2\right)\nonumber\\
&&+\left(|H_{\frac{1}{2},t}^P|^2+|H_{-\frac{1}{2},t}^P|^2\right)\cdot\left(2q^2-3m_\ell^2\right)-8m_\ell\sqrt{q^2}\cdot{\rm Re}\left[H_{\frac{1}{2},t}^{A*}H_{\frac{1}{2},t}^P+H_{-\frac{1}{2},t}^{A*}H_{-\frac{1}{2},t}^P\right]\nonumber\\
&&+\left(|H_{\frac{1}{2},0}^V|^2+|H_{-\frac{1}{2},0}^V|^2\right)\cdot\left(8m_\ell^2+2q^2\beta^2\right)
+\left(|H_{\frac{1}{2},0}^A|^2+|H_{-\frac{1}{2},0}^A|^2\right)\cdot 2q^2\beta^2\nonumber\\
&&+\left(|H_{\frac{1}{2},+}^V|^2+|H_{-\frac{1}{2},-}^V|^2\right)\cdot\left(8m_\ell^2+q^2\beta^2\right)
+\left(|H_{\frac{1}{2},+}^A|^2+|H_{-\frac{1}{2},-}^A|^2\right)\cdot q^2\beta^2,\nonumber\\
I_2(q^2)=&&-8m_\ell\sqrt{q^2}\beta\cdot{\rm Re}\left[H_{\frac{1}{2},t}^SH_{\frac{1}{2},0}^{V*}+H_{-\frac{1}{2},t}^SH_{-\frac{1}{2},0}^{V*}\right]-4q^2\beta\cdot{\rm Re}\left[H_{\frac{1}{2},+}^AH_{\frac{1}{2},+}^{V*}-H_{-\frac{1}{2},-}^AH_{-\frac{1}{2},-}^{V*}\right],\nonumber\\
I_3(q^2)=&&\left(|H_{\frac{1}{2},0}^V|^2
+|H_{-\frac{1}{2},0}^V|^2+|H_{\frac{1}{2},0}^A|^2+|H_{-\frac{1}{2},0}^A|^2\right)\cdot\left(-2q^2\beta^2\right)\nonumber\\
&&+\left(|H_{\frac{1}{2},+}^V|^2+|H_{-\frac{1}{2},-}^V|^2+|H_{\frac{1}{2},+}^A|^2+|H_{-\frac{1}{2},-}^A|^2\right)\cdot q^2\beta^2,\label{eq:B1tpB2mumuUnbinnedCC}
\end{eqnarray}
where $q=p_1-p_2$, ${\cal N}_f(q^2)=\frac{G_F^2\alpha^2|\lambda_t|^2\beta_\ell}{2048\pi^3}\frac{\sqrt{Q_+Q_-}}{M_1^3}$, $Q_\pm=(M_1\pm M_2)^2-q^2$, and $\beta_\ell=\sqrt{1-\frac{4m_\ell^2}{q^2}}$. The $\theta_\ell$ is the angle between the three-momenta of the final state particle $B_2$ and the emitting direction of $\ell^-$. The hadronic amplitudes in Eq.~(\ref{eq:B1tpB2mumuUnbinnedCC}) are given as follows:
{\small
\begin{align}
&H_{\frac{1}{2},t}^V=\frac{\sqrt{Q_+}\left(C_9+C_9'\right)\left[f_1(q^2)(M_1-M_2)+f_3(q^2)\frac{q^2}{M_1}\right]}
{\sqrt{q^2}}+\frac{\sqrt{Q_-}\left(C_9-C_9'\right)\left[g_1(q^2)(M_1+M_2)-g_3(q^2)\frac{q^2}{M_1}\right]}
{\sqrt{q^2}}\nonumber\\
&\qquad\qquad-\frac{4\sqrt{2}\pi\left[c\cdot\sqrt{Q_+}(M_1-M_2)-d\cdot\sqrt{Q_-}(M_1+M_2)\right]}
{\sqrt{q^2}\lambda_t},\nonumber\\
&H_{-\frac{1}{2},t}^V=\frac{\sqrt{Q_+}\left(C_9+C_9'\right)\left[f_1(q^2)(M_1-M_2)+f_3(q^2)\frac{q^2}{M_1}\right]}
{\sqrt{q^2}}-\frac{\sqrt{Q_-}\left(C_9-C_9'\right)\left[g_1(q^2)(M_1+M_2)-g_3(q^2)\frac{q^2}{M_1}\right]}
{\sqrt{q^2}}\nonumber\\
&\qquad\qquad-\frac{4\sqrt{2}\pi\left[c\cdot\sqrt{Q_+}(M_1-M_2)+d\cdot\sqrt{Q_-}(M_1+M_2)\right]}
{\sqrt{q^2}\lambda_t},\nonumber\\
&H_{\frac{1}{2},0}^V=\frac{ C_7m_s\left[\sqrt{Q_-}\left(g_T(q^2)-g_T^{(1)}(q^2)\frac{M_1+M_2}{M_1}-g_T^{(2)}(q^2)\frac{Q_+}{M_1^2}\right)
+\sqrt{Q_+}\left(g_{T5}(q^2)+g_{T5}^{(1)}(q^2)\frac{M_1-M_2}{M_1}-g_T^{(2)}(q^2)\frac{Q_-}{M_1^2}\right)\right]}{2\pi\sqrt{q^2}}\nonumber\\
&\qquad\qquad+\frac{\sqrt{Q_-}\left(C_9+C_9'\right)\left[f_1(q^2)(M_1+M_2)+f_2(q^2)\frac{q^2}{M_1}\right]}{\sqrt{q^2}}
+\frac{\sqrt{Q_+}\left(C_9-C_9'\right)\left[g_1(q^2)(M_1-M_2)-g_2(q^2)\frac{q^2}{M_1}\right]}{\sqrt{q^2}}\nonumber\\
&\qquad\qquad+\frac{4\sqrt{2}\pi[-a\sqrt{Q_-}-b\sqrt{Q_+}-c\sqrt{Q_-}(M_1+M_2)+d\sqrt{Q_+}(M_1-M_2)]}
{\sqrt{q^2}\lambda_t},\nonumber\\
&H_{-\frac{1}{2},0}^V=\frac{ C_7m_s\left[\sqrt{Q_-}\left(g_T(q^2)-g_T^{(1)}(q^2)\frac{M_1+M_2}{M_1}-g_T^{(2)}(q^2)\frac{Q_+}{M_1^2}\right)
-\sqrt{Q_+}\left(g_{T5}(q^2)+g_{T5}^{(1)}(q^2)\frac{M_1-M_2}{M_1}-g_T^{(2)}(q^2)\frac{Q_-}{M_1^2}\right)\right]}{2\pi\sqrt{q^2}}\nonumber\\
&\qquad\qquad+\frac{\sqrt{Q_-}\left(C_9+C_9'\right)\left[f_1(q^2)(M_1+M_2)+f_2(q^2)\frac{q^2}{M_1}\right]}{\sqrt{q^2}}
-\frac{\sqrt{Q_+}\left(C_9-C_9'\right)\left[g_1(q^2)(M_1-M_2)-g_2(q^2)\frac{q^2}{M_1}\right]}{\sqrt{q^2}}\nonumber\\
&\qquad\qquad+\frac{4\sqrt{2}\pi[-a\sqrt{Q_-}+b\sqrt{Q_+}-c\sqrt{Q_-}(M_1+M_2)-d\sqrt{Q_+}(M_1-M_2)]}
{\sqrt{q^2}\lambda_t},\nonumber\\
&H_{\frac{1}{2},+}^V=-\frac{C_7m_s\left[g_T(q^2)\sqrt{Q_-}(M_1+M_2)-g_T^{(1)}(q^2)\sqrt{Q_-}\frac{q^2}{M_1}+g_{T5}(q^2)\sqrt{Q_+}(M_1-M_2)
+g_{T5}^{(1)}(q^2)\sqrt{Q_+}\frac{q^2}{M_1}\right]}{2\pi q^2}\nonumber\\
&\qquad\qquad-\sqrt{2Q_-}\left(C_9+C_9'\right)\left[f_1(q^2)+f_2(q^2)\frac{M_1+M_2}{M_1}\right]
-\sqrt{2Q_+}\left(C_9-C_9'\right)\left[g_1(q^2)-g_2(q^2)\frac{M_1-M_2}{M_1}\right]\nonumber\\
&\qquad\qquad-\frac{8\pi\left[-a\sqrt{Q_-}(M_1+M_2)-b\sqrt{Q_+}(M_1-M_2)-c\sqrt{Q_-}q^2+d\sqrt{Q_+}q^2\right]}
{q^2\lambda_t},\nonumber\\
&H_{-\frac{1}{2},-}^V=-\frac{C_7m_s\left[g_T(q^2)\sqrt{Q_-}(M_1+M_2)-g_T^{(1)}(q^2)\sqrt{Q_-}\frac{q^2}{M_1}-g_{T5}(q^2)\sqrt{Q_+}(M_1-M_2)
-g_{T5}^{(1)}(q^2)\sqrt{Q_+}\frac{q^2}{M_1}\right]}{2\pi q^2}\nonumber\\
&\qquad\qquad-\sqrt{2Q_-}\left(C_9+C_9'\right)\left[f_1(q^2)+f_2(q^2)\frac{M_1+M_2}{M_1}\right]
+\sqrt{2Q_+}\left(C_9-C_9'\right)\left[g_1(q^2)-g_2(q^2)\frac{M_1-M_2}{M_1}\right]\nonumber\\
&\qquad\qquad-\frac{8\pi\left[-a\sqrt{Q_-}(M_1+M_2)+b\sqrt{Q_+}(M_1-M_2)-c\sqrt{Q_-}q^2-d\sqrt{Q_+}q^2\right]}
{q^2\lambda_t},\nonumber\\
&H_{\frac{1}{2},t}^A=\frac{\sqrt{Q_+}\left(C_{10}+C_{10}'\right)\left[f_1(q^2)(M_1-M_2)+f_2(q^2)\frac{q^2}{M_1}\right]}
{\sqrt{q^2}}+\frac{\sqrt{Q_-}\left(C_{10}-C_{10}'\right)\left[g_1(q^2)(M_1+M_2)-g_2(q^2)\frac{q^2}{M_1}\right]}
{\sqrt{q^2}},\nonumber\\
&H_{-\frac{1}{2},t}^A=\frac{\sqrt{Q_+}\left(C_{10}+C_{10}'\right)\left[f_1(q^2)(M_1-M_2)+f_2(q^2)\frac{q^2}{M_1}\right]}
{\sqrt{q^2}}-\frac{\sqrt{Q_-}\left(C_{10}-C_{10}'\right)\left[g_1(q^2)(M_1+M_2)-g_2(q^2)\frac{q^2}{M_1}\right]}
{\sqrt{q^2}},\nonumber\\
&H_{\frac{1}{2},0}^A=\frac{\sqrt{Q_-}\left(C_{10}+C_{10}'\right)\left[f_1(q^2)(M_1+M_2)+f_2(q^2)\frac{q^2}{M_1}\right]}
{\sqrt{q^2}}
+\frac{\sqrt{Q_+}\left(C_{10}-C_{10}'\right)\left[g_1(q^2)(M_1-M_2)-g_2(q^2)\frac{q^2}{M_1}\right]}
{\sqrt{q^2}},\nonumber\\
&H_{-\frac{1}{2},0}^A=\frac{\sqrt{Q_-}\left(C_{10}+C_{10}'\right)\left[f_1(q^2)(M_1+M_2)+f_2(q^2)\frac{q^2}{M_1}\right]}
{\sqrt{q^2}}
-\frac{\sqrt{Q_+}\left(C_{10}-C_{10}'\right)\left[g_1(q^2)(M_1-M_2)-g_2(q^2)\frac{q^2}{M_1}\right]}
{\sqrt{q^2}},\nonumber\\
&H_{\frac{1}{2},1}^A=-\sqrt{2Q_-}\left(C_{10}+C_{10}'\right)\left[f_1(q^2)+f_2(q^2)\frac{M_1+M_2}{M_1}\right]
-\sqrt{2Q_+}\left(C_{10}-C_{10}'\right)\left[g_1(q^2)-g_2(q^2)\frac{M_1-M_2}{M_1}\right],\nonumber\\
&H_{-\frac{1}{2},-1}^A=-\sqrt{2Q_-}\left(C_{10}+C_{10}'\right)\left[f_1(q^2)+f_2(q^2)\frac{M_1+M_2}{M_1}\right]
+\sqrt{2Q_+}\left(C_{10}-C_{10}'\right)\left[g_1(q^2)-g_2(q^2)\frac{M_1-M_2}{M_1}\right],\nonumber\\
&H_{\frac{1}{2},t}^S=\sqrt{Q_+}\left(C_S+C_S'\right)f_S+\sqrt{Q_-}\left(C_S-C_S'\right)g_P,\nonumber\\
&H_{-\frac{1}{2},t}^S=\sqrt{Q_+}\left(C_S+C_S'\right)f_S-\sqrt{Q_-}\left(C_S-C_S'\right)g_P,\nonumber\\
&H_{\frac{1}{2},t}^P=\sqrt{Q_+}\left(C_P+C_P'\right)f_S+\sqrt{Q_-}\left(C_P-C_P'\right)g_P,\nonumber\\
&H_{-\frac{1}{2},t}^P=\sqrt{Q_+}\left(C_P+C_P'\right)f_S-\sqrt{Q_-}\left(C_P-C_P'\right)g_P.
\end{align}
}
\section{Renormalization group evolution of the SM dimension-six operator coefficients}

The one-loop RG equations of the dimension-six operator coefficients are:
\begin{eqnarray}
16\pi^2\mu\frac{dC_i(\mu)}{d\mu}=\gamma_{ij}C_j(\Lambda)\equiv\mbox{\.{C}}_i(\Lambda),
\end{eqnarray}
where $\mbox{\.{C}}_i(\Lambda)$ can be found in Refs.~\cite{Jenkins:2013zja,Jenkins:2013wua,Alonso:2013hga}. Therein, these coefficients $C_{Hq}^{(1)}$, $C_{H\ell}^{(1)}$, $C_{\ell\ell}$, $C_{qq}^{(1)}$, $C_{qq}^{(3)}$, $C_{qu}^{(1)}$ and $C_{qd}^{(1)}$ do not contribute to semi-leptonic decays. Furthermore, the indices $ww$ in the terms like $\left[\cdots\right]_{ww\alpha\beta}\delta_{ij}$ denote two identical quarks. These terms do not contribute to FCNC transitions. Therefore, the $\mbox{\.{C}}_i(\Lambda)$ contributing to down-quark FCNC semi-leptonic decays can be written as:
\begin{eqnarray}
&&\left[\mbox{\.{C}}_{\ell q}^{(1)}(\Lambda)\right]_{ij\alpha\beta}=\frac{8}{3}g_1^2y_\ell^2\left[C_{\ell q}^{(1)}(\Lambda)\right]_{ijww}\delta_{\alpha\beta}+\frac{4}{3}g_1^2y_ey_\ell\left[C_{qe}(\Lambda)\right]_{ijww}\delta_{\alpha\beta}\nonumber\\
&&\qquad\qquad\qquad+12g_1^2y_\ell y_q\left[C_{\ell q}^{(1)}(\Lambda)\right]_{ij\alpha\beta}
+9g_2^2\left[C_{\ell q}^{(3)}(\Lambda)\right]_{ij\alpha\beta},\nonumber\\
&&\left[\mbox{\.{C}}_{\ell q}^{(3)}(\Lambda)\right]_{ij\alpha\beta}=\frac{2}{3}g_2^2\left[C_{\ell q}^{(3)}(\Lambda)\right]_{ijww}\delta_{\alpha\beta}
+3g_2^2\left[C_{\ell q}^{(1)}(\Lambda)\right]_{ij\alpha\beta}-6\left(g_2^2-2y_\ell y_qg_1^2\right)\left[C_{\ell q}^{(3)}(\Lambda)\right]_{ij\alpha\beta},\nonumber\\
&&\left[\mbox{\.{C}}_{\ell d}(\Lambda)\right]_{ij\alpha\beta}=\frac{8}{3}g_1^2y_\ell^2\left[C_{\ell d}(\Lambda)\right]_{ijww}\delta_{\alpha\beta}
+\frac{4}{3}g_1^2y_ey_\ell\left[C_{ed}(\Lambda)\right]_{ijww}\delta_{\alpha\beta}-12y_\ell y_dg_1^2\left[C_{\ell d}(\Lambda)\right]_{ij\alpha\beta},\nonumber\\
&&\left[\mbox{\.{C}}_{qe}(\Lambda)\right]_{ij\alpha\beta}=\frac{8}{3}g_1^2N_cy_q^2\left[C_{qe}(\Lambda)\right]_{ijww}\delta_{\alpha\beta}
+\frac{4}{3}g_1^2N_cy_qy_d\left[C_{ed}(\Lambda)\right]_{ijww}\delta_{\alpha\beta}-12y_qy_eg_1^2\left[C_{qe}(\Lambda)\right]_{ij\alpha\beta},\nonumber\\
&&\left[\mbox{\.{C}}_{ed}(\Lambda)\right]_{ij\alpha\beta}=\frac{8}{3}g_1^2y_ey_\ell\left[C_{\ell d}(\Lambda)\right]_{ijww}\delta_{\alpha\beta}
+\frac{4}{3}g_1^2y_e^2\left[C_{ed}(\Lambda)\right]_{ijww}\delta_{\alpha\beta}+12y_ey_dg_1^2\left[C_{ed}(\Lambda)\right]_{ij\alpha\beta},\nonumber\\
&&\left[\mbox{\.{C}}_{\ell edq}(\Lambda)\right]_{ij\alpha\beta}=-\left(6\left(y_d\left(y_q-y_e\right)+y_e\left(y_e+y_q\right)\right)g_1^2+3\left(N_c-\frac{1}{N_c}\right)g_3^2\right)
\left[C_{\ell edq}(\Lambda)\right]_{ij\alpha\beta},
\end{eqnarray}
where $y_q=\frac{1}{6}$, $y_d=-\frac{1}{3}$, $y_\ell=-\frac{1}{2}$ and $y_e=-1$ are hypercharge. Three gauge couplings $g_1$, $g_2$ and $g_3$ are defined as $g_1=\frac{e}{\cos\theta_W}=0.36$, $g_2=\frac{e}{\sin\theta_W}=0.65$ and $g_3=\sqrt{4\pi\alpha_s}=1.22$, respectively, where $\alpha_s$ is the strong coupling constant and $N_c=3$ is the number of colors.

When running the RG equations from the NP scale $\Lambda$ to the electroweak scale $\upsilon$, the dimension-six operator coefficients for the case of $\alpha=\beta$ read:
{\small
\begin{eqnarray}
&&\left[C_{\ell q}^{(1)}(\upsilon)\right]_{ij\alpha\alpha}=\left[1-\frac{1}{16\pi^2}\left(\frac{8}{3}g_1^2y_\ell^2+12g_1^2y_\ell y_q\right){\rm In}\frac{\Lambda}{\upsilon}\right]\left[C_{\ell q}^{(1)}(\Lambda)\right]_{ij\alpha\alpha}\nonumber\\
&&\qquad\qquad\qquad+\left[-\frac{9}{16\pi^2}g_2^2{\rm In}\frac{\Lambda}{\upsilon}\right]\left[C_{\ell q}^{(3)}(\Lambda)\right]_{ij\alpha\alpha}+\left[-\frac{1}{12\pi^2}g_1^2y_ey_\ell{\rm In}\frac{\Lambda}{\upsilon}\right]\left[C_{qe}(\Lambda)\right]_{ij\alpha\alpha},\nonumber\\
&&\left[C_{\ell q}^{(3)}(\upsilon)\right]_{ij\alpha\alpha}=\left[1-\frac{1}{16\pi^2}\left(-\frac{16}{3}g_2^2+12y_\ell y_qg_1^2\right){\rm In}\frac{\Lambda}{\upsilon}\right]\left[C_{\ell q}^{(3)}(\Lambda)\right]_{ij\alpha\alpha}+\left[-\frac{3}{16\pi^2}g_2^2{\rm In}\frac{\Lambda}{\upsilon}\right]\left[C_{\ell q}^{(1)}(\Lambda)\right]_{ij\alpha\alpha},\nonumber\\
&&\left[C_{\ell d}(\upsilon)\right]_{ij\alpha\alpha}=\left[1-\frac{1}{16\pi^2}\left(\frac{8}{3}g_1^2y_\ell^2-12y_\ell y_dg_1^2\right){\rm In}\frac{\Lambda}{\upsilon}\right]\left[C_{\ell d}(\Lambda)\right]_{ij\alpha\alpha}+\left[-\frac{1}{12\pi^2}g_1^2y_ey_\ell{\rm In}\frac{\Lambda}{\upsilon}\right]\left[C_{ed}(\Lambda)\right]_{ij\alpha\alpha},\nonumber\\
&&\left[C_{qe}(\upsilon)\right]_{ij\alpha\alpha}=\left[1-\frac{1}{16\pi^2}\left(\frac{8}{3}g_1^2N_cy_q^2-12y_q y_eg_1^2\right){\rm In}\frac{\Lambda}{\upsilon}\right]\left[C_{qe}(\Lambda)\right]_{ij\alpha\alpha}\nonumber\\
&&\qquad\qquad\qquad+\left[-\frac{1}{12\pi^2}g_1^2N_cy_qy_d{\rm In}\frac{\Lambda}{\upsilon}\right]\left[C_{ed}(\Lambda)\right]_{ij\alpha\alpha},\nonumber\\
&&\left[C_{ed}(\upsilon)\right]_{ij\alpha\alpha}=\left[1-\frac{1}{16\pi^2}\left(\frac{4}{3}g_1^2y_e^2+12y_e y_dg_1^2\right){\rm In}\frac{\Lambda}{\upsilon}\right]\left[C_{ed}(\Lambda)\right]_{ij\alpha\alpha}+\left[-\frac{1}{6\pi^2}g_1^2y_ey_\ell{\rm In}\frac{\Lambda}{\upsilon}\right]\left[C_{\ell d}(\Lambda)\right]_{ij\alpha\alpha},\nonumber\\
&&\left[C_{\ell edq}(\upsilon)\right]_{ij\alpha\alpha}=\left[1+\frac{1}{16\pi^2}\left(6\left(y_d\left(y_q-y_e\right)+y_e\left(y_e+y_q\right)\right)g_1^2+3\left(N_c-\frac{1}{N_c}\right)g_3^2\right){\rm In}\frac{\Lambda}{\upsilon}\right]\left[C_{\ell edq}(\Lambda)\right]_{ij\alpha\alpha}.\nonumber\\
\end{eqnarray}
}


\end{document}